\documentclass{ws-procs11x85}
\usepackage{ws-procs-thm}
\usepackage{multirow}
\usepackage{array}
\usepackage{tabularray}

\usepackage{etoolbox}
\newtoggle{showtodo}
\toggletrue{showtodo}

\begin{document}
\title{Graph algorithms for predicting subcellular localization at the pathway level}

\author{Chris S Magnano$^{1,2,3}$ and Anthony Gitter$^{1,2,4}$}

\address{$^{1}$Department of Computer Sciences, University of Wisconsin-Madison, Madison, WI, USA \\
$^{2}$Morgridge Institute for Research, Madison, WI, USA \\
$^{3}$Center for Computational Biomedicine, Harvard Medical School, Boston, MA, USA\\
$^{4}$Department of Biostatistics and Medical Informatics, University of Wisconsin-Madison, Madison, WI, USA}

\begin{abstract}
Protein subcellular localization is an important factor in normal cellular processes and disease.
While many protein localization resources treat it as static, protein localization is dynamic and heavily influenced by biological context. 
Biological pathways are graphs that represent a specific biological context and can be inferred from large-scale data. 
We develop graph algorithms to predict the localization of all interactions in a biological pathway as an edge-labeling task. 
We compare a variety of models including graph neural networks, probabilistic graphical models, and discriminative classifiers for predicting localization annotations from curated pathway databases.
We also perform a case study where we construct biological pathways and predict localizations of human fibroblasts undergoing viral infection.
Pathway localization prediction is a promising approach for integrating publicly available localization data into the analysis of large-scale biological data. 
\end{abstract}


% required, do-not-remove (do remove for preprint)
%\copyrightinfo{\copyright\ 2022 The Authors. Open Access chapter published by World Scientific Publishing Company and distributed under the terms of the Creative Commons Attribution Non-Commercial (CC BY-NC) 4.0 License.}
% Reuse the \copyrightinfo for the preprint for the software because the command is important for the footnote nubmering
\copyrightinfo{Code can be found at \url{https://github.com/gitter-lab/pathway-localization} and is archived at \url{https://doi.org/10.5281/zenodo.7140733}}

\section{Introduction}
%%% Introduce cell state and localization
Cellular state is dictated by a wide range of factors from chromatin accessibility to protein abundance to the physical location of proteins within the cell.
%Bringing additional sources of information to bear in biological analyses can elucidate these factors role and provide explanation for biological phenomena. 
Cells are compartmentalized into subcellular locations that provide the chemical environment around proteins.
That local environment informs proteins' structure and available interaction partners.  
%How proteins localize to subcellular compartments is key information for understanding biological processes~\cite{lundbergSpatialProteomicsPowerful2019}. 
Protein localization not only dictates protein interactions in normal biological processes~\cite{lundbergSpatialProteomicsPowerful2019}, but also is an important factor that can contribute to abnormal cellular behavior. 
Alzheimer's disease, amyotrophic lateral sclerosis, Wilson disease, and multiple cancers involve abnormal protein localizations~\cite{hung_protein_2011}. 
%Therefore, knowing these non-standard localization can give insight into how a disease operates at a cellular level and which proteins may be noteworthy. 

%%% Localization is actually context specific and dynamic, but most resources and models are static
Although protein localization is dynamic and context-specific~\cite{mechanismLocalizationReview}, many localization resources present a fixed, static view.
Localization databases such as MatrixDB~\cite{chautard_matrixdb_2010}, Organelle DB~\cite{wiwatwattana_organelle_2005}, Compartments~\cite{binder_compartments_2014}, and ComPPI~\cite{veres_comppi_2015} track primary experimental data, computational predictions, or combinations of multiple information sources.
Up to $50\%$ of proteins localize to multiple cellular compartments~\cite{thul_subcellular_2017, zhangDBMLocDatabaseProteins2008}.
Databases typically provide multiple possible localizations per protein, but that does not determine the conditions under which subsets of each protein's localiziations are relevant.
%If trying to apply this localization information to a particular experimental context, there is no systematic way to choose for every protein of interest what subset of its possible localizations are applicable. 
%Furthermore proteins do not localize independently of other proteins, so the multiple possible localizations for all proteins of interest must be considered simultaneously.
Many tools can predict possible locations of a protein based on its sequence~\cite{gardy_methods_2006, imaiPredictionSubcellularLocations2010, alaaProteinSubcellularLocalization2019} using machine learning methods such as logistic regression~\cite{hua_support_2001} or deep neural networks~\cite{almagro_armenteros_deeploc_2017}. % this is a big area, may want to reference a few additional landmark papers or give a sense of why sequence carries any information about localization
Some methods incorporate additional information, such as gene expression~\cite{drawid_bayesian_2000}, Gene Ontology annotations~\cite{fyshe_improving_2008}, and network information~\cite{anandaNetLocNetworkBased2010, duPredictingHumanProtein2014, garapatiPredictingSubcellularLocalization2020, grover_protFinder}.
Methods using network information consider the localizations of neighboring proteins in protein-protein interaction databases to aid in localization prediction and do not attempt to represent any particular biological context. % the class of method using network information is the most relevant to this work, so it would be nice to expand on that in a new paragraph providing ~1 sentence per method; also may need to consider one or more of those methods (or methods inspired by them, like node2vec) as baselines
%However, localization data from databases or computational predictions cannot be easily integrated with other types of biological data that are context-specific. 
%Though predictive methods tend to consider protein localization as static, localization in reality is a dynamic process mediated by a variety of physical factors and biological processes~\cite{mechanismLocalizationReview}.
Some predictive methods consider tissue context~\cite{zhuTissueSpecificSubcellularLocalization2019}, but proteins vary in their subcellular localization even between single cells of the same tissue type~\cite{lundbergSpatialProteomicsPowerful2019}.

%%% Graph algorithms can help estimate context specific localization, pathway reconstruction + our new annotation approach
We present graph algorithms for estimating context-specific protein localizations by modeling them in biological pathways. %\footnote{Supplementary Information and code can be found at \url{https://github.com/gitter-lab/pathway-localization} and archived at \url{https://doi.org/10.5281/zenodo.7140733}.}.
% Move this earlier in the manuscript for the preprint
Biological pathways, graphs of biological entities such as proteins, can represent a particular biological process or context.
Although traditionally thought of in terms of curated pathway databases, pathway reconstruction graph algorithms~\cite{pathLinker,omicsIntegrator,netbox} can generate custom pathway representations of a specific process given a background protein interaction network and condition-specific data such as proteomic measurements as input. % Add ResponseNet to the list of space permits
However, there is no straightforward way to contextualize and apply available protein localization data to this type of predicted biological pathway.
%Pathways created via pathway reconstruction represent the context of whatever data was used to construct them.
%This allows a pathway to be created for whatever data is being studies, and circumvents the need for a curated pathway to already exist in a database.
%Thus, there is a need to provide context-specific localization information that accounts for cellular state.  
In order to provide context-specific localization information for a particular biological dataset, we develop graph algorithms for the simultaneous prediction of a subcellular localization for all interactions in a reconstructed biological pathway. 
Computationally, this can be seen as an edge labeling task on an existing graph.
This predictive step can be added to existing pathway reconstruction workflows.
Estimating localization information at the pathway level enables examining where proteins or other biological entities are when they perform a biological function. %, and allow for a clear application of localization information to a particular cellular state.
Pathway-specific localization annotation can help interpret the predicted pathway and potentially provide additional information to guide followup experiments.

%%% advantages over pathway databases, experiments
Our strategy to understand context-specific protein localization through graph-based annotations of reconstructed pathways offers advantages over alternative approaches.
Some curated pathway databases provide localization information at the interaction level and include information about non-protein biological entities~\cite{reactome, pathBank}. 
However, many pathway databases contain incomplete or no localization information.
For instance, of the $8$ pathway databases included in Pathway Commons~\cite{pathwayCommons}, $2$ are fully labeled with localization information, $5$ are partially labeled with localization information, and $1$ contains no programmatically available localization information.
Additionally, curated pathways often do not line up with experimental data~\cite{TPS, tCellSignalingPhos2012, insulinPhos2015a, tgfBetaPhos2014} and a curated pathway may not be available for a particular biological condition of interest. % Remove some example references here if space is needed
While condition-specific localization information can be experimentally derived~\cite{lundbergSpatialProteomicsPowerful2019} using mass spectrometry or cellular imaging, these methods can be expensive, require experimental expertise, and have incomplete coverage.
Predicting localization based on pathways is less precise than acquiring localization data experimentally, but the predictions  provide an initial coarse estimate of all proteins' localizations without requiring new specialized data.
%Ideally, localization information could be added to an existing analysis without performing an additional experiment. 
%However, current localization data sources do not provide this information in specific functional contexts or in a consistent, complete way.

We develop and compare three categories of methods for predicting localization for interactions within the context of a biological pathway: graph neural networks, probabilistic graphical models, and classifiers that do not use graph topology. 
First, we quantitatively evaluate these strategies for pathway-based localization prediction by holding out annotated localizations from pathway databases.
Then, we demonstrate how our approach can be used in practice with a case study involving human cytomegalovirus (HCMV) infection over time~\cite{jeanbeltranPortraitHumanOrganelle2016}.
While there are disparities between localization information in pathway databases and experimentally-derived localization data, pathway-level localization prediction is a promising approach for combining publicly available localization data with the analysis of large-scale biological data. % Let's keep thinking about our overall message and conclusion

\section{Methods}
\subsection{Pathway Localization Prediction Problem Definition}

\label{sec:methods_locPred}

    \begin{figure*}[hbtp]
         \centering
         \includegraphics[width=\textwidth]{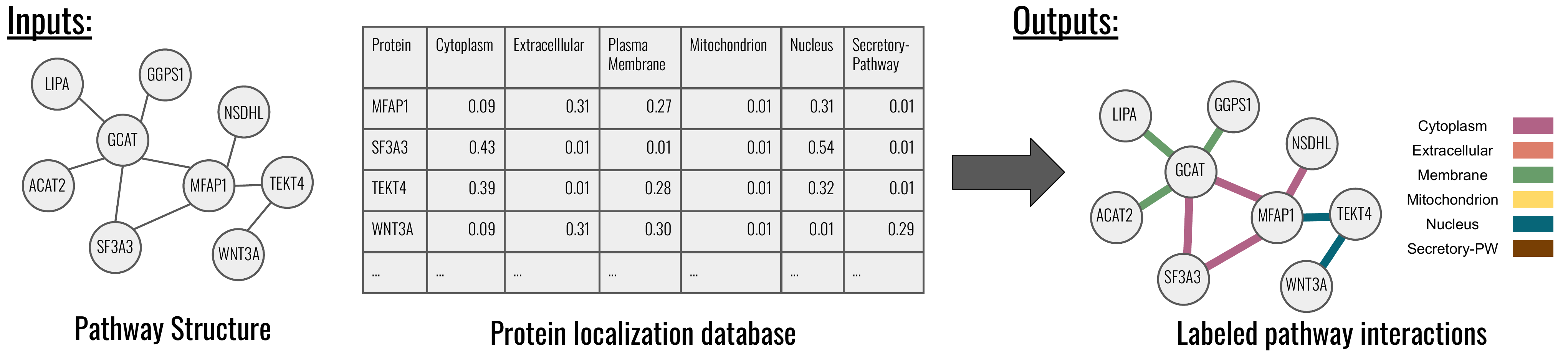}
         \caption{Overview of the pathway localization prediction experimental workflow.}
         \label{fig:loc_overview}
     \end{figure*} % Make the 'S' in Pathway Structure lower case
     % Could be more descriptive about the inputs and outputs here

Given a biological pathway represented as a graph, the goal is to predict one subcellular localization for each edge. % Save the formal notation for below
The pathway represents some cellular function and can be constructed from large-scale biological datasets using pathway reconstruction\cite{PPA}. 
We predict a localization for each edge in the pathway, which can be viewed as a class label assignment for each edge in the graph. 
Protein-level localization information is used as input to the prediction task as node features. 
Thus, the pathway-specific subcellular localization task can be defined as:

\underline{Input:} (1) A context-specific pathway graph consisting of nodes and edges $G = (N, E)$, and (2) a distribution over possible localizations for each node in the graph.  % Note that the graph is undirected?
\underline{Output:} A single localization assignment for each interaction $e \in E$. See Figure~\ref{fig:loc_overview}.

We chose to assign localizations to edges as opposed to nodes and to assign each interaction a single localization. 
Pathway databases such as Reactome~\cite{reactome} and popular pathway file formats such as BioPax~\cite{demirBioPAXCommunityStandard2010,gyori_pybiopax_2022} only allow proteins to be in a single subcellular location, creating multiple protein entries if they occur in multiple localizations and assigning them to interactions. 
While many proteins have multiple localizations, among all Reactome and PathBank pathways less than $5\%$ of total interactions have multiple localizations within the same pathway. 

\subsection{Experimental Setup}

\subsubsection{Pathway Database Localization Prediction}
\label{sec:exp_path_data}
We investigated how well protein localization databases can be used to predict context-specific localizations in pathway databases, both to examine the feasibility of pathway-specific localization prediction and to elucidate the relationship between node labels in protein localization databases and edge labels in pathway databases. 
Pathways with interaction localization labels from the Reactome~\cite{reactome} and PathBank~\cite{pathBank} databases were each used as ground truth datasets. 

The original pathways in both Reactome and PathBank are represented as hypergraphs, where reaction edges can contain more than two nodes. 
Pathway Commons converts these hypergraphs to graphs using a set of rules\footnote{\url{http://www.pathwaycommons.org/pc2/formats}}.
To represent a protein-complex that contains $n$ proteins, the hypergraph conversions create an edge between every possible pair of nodes, resulting in $n^2$ edges.
For instance, the 4 hyperedges that make up the PathBank pathway Protein Synthesis: Serine are converted to 3,318 edges, of which 3,315 are of type ``in-complex-with''.
We collapsed protein complexes into single nodes where possible in all pathways. 
This was done by removing any nodes if all of its edges were redundant with the protein-complex's edges, leaving a single node for each complex.
Though this loses some node information, collapsing protein complexes resulted in pathways that more more closely resembled the original hypergraph in edge distribution, topology, and class balance.

Three different node feature sets were used: the ComPPI database~\cite{veres_comppi_2015}, the Compartments database~\cite{binder_compartments_2014}, and UniProt keyword\cite{uniprot} features. 
ComPPI and Compartments contain localization scores for each protein, which are used directly as input features. 
We created a dimensionality reduction-based vectorization of UniProt keyword assignments for all proteins (Section \ref{sec:uniprot_kw_feats}).
All $8$ predictive models (Section~\ref{sec:main_models}) were tested on all feature sets with the exception of the NaivePGM model, which could not use the UniProt keyword features as it interprets input features directly as conditional probabilities.
All pathways in the 2 pathway databases Reactome and PathBank, which contain interaction-level localization labels, were tested on resulting in a total of $46$ runs.
Models were trained using 5-fold cross validation, and model selection and hyperparameter selection were performed on a tuning set of the $53$ Reactome pathways categorized as developmental and a randomly chosen $10\%$ of all PathBank pathways.
Tuning pathways were excluded from cross validation.

\subsubsection{Human Cytomegalovirus Case Study}
To examine how predicting context-specific localization at the pathway level could be used in a realistic setting, we performed a case study with bulk spatial mass spectrometry (MS) data from multi-organelle profiling on primary fibroblasts during HCMV infection~\cite{jeanbeltranPortraitHumanOrganelle2016}.
In multi-organelle profiling, gradient centrifugation is used on a bulk sample to partially separate organelles. 
Protein levels in each subcellular fraction are then measured using tandem mass tags MS, and localization labels are determined by clustering proteins with similar fraction profiles.
We investigated whether a predictive model can infer localizations in the context of viral infection, potentially bypassing the need to collect spatial proteomic data. 

We performed pathway reconstruction~\cite{PPA} by combining a background protein-protein interaction network~\cite{TPS,irefindex,phosphosite} with label-free MS data, which measured protein abundance across the entire fibroblast at $120$ hours post infection (hpi) without regards to localization.
Measured protein levels were used to create biological networks representing the cell state following infection.
The combined top pathways chosen (Section~\ref{sec:methods_casestudy}) contained a total of $386$ edges with localization information at 120hpi. 

We then trained one of the best performing models from the pathway database prediction task, the graph attention network, in three different scenarios. 
First, we trained a model using data from the PathBank database as described in Section \ref{sec:methods_locPred}. 
Second, we trained a model using a separate dataset that measured protein localization using a similar method on a different cell type and under a different biological condition, HeLa cells undergoing EGF stimulation~\cite{hela_proteomics}. 
Third, we trained a model on the same HCMV experiment at the 24hpi timepoint. 
This third scenario is unlikely to occur, as it would require a dataset to already exist for an identical cell type and condition, but gives a useful benchmark for best case predictive performance. 

\subsection{Pathway Localization Prediction Models}
\label{sec:main_models}
\begin{figure*}[htbp]
         \centering
         \includegraphics[width=0.7\textwidth]{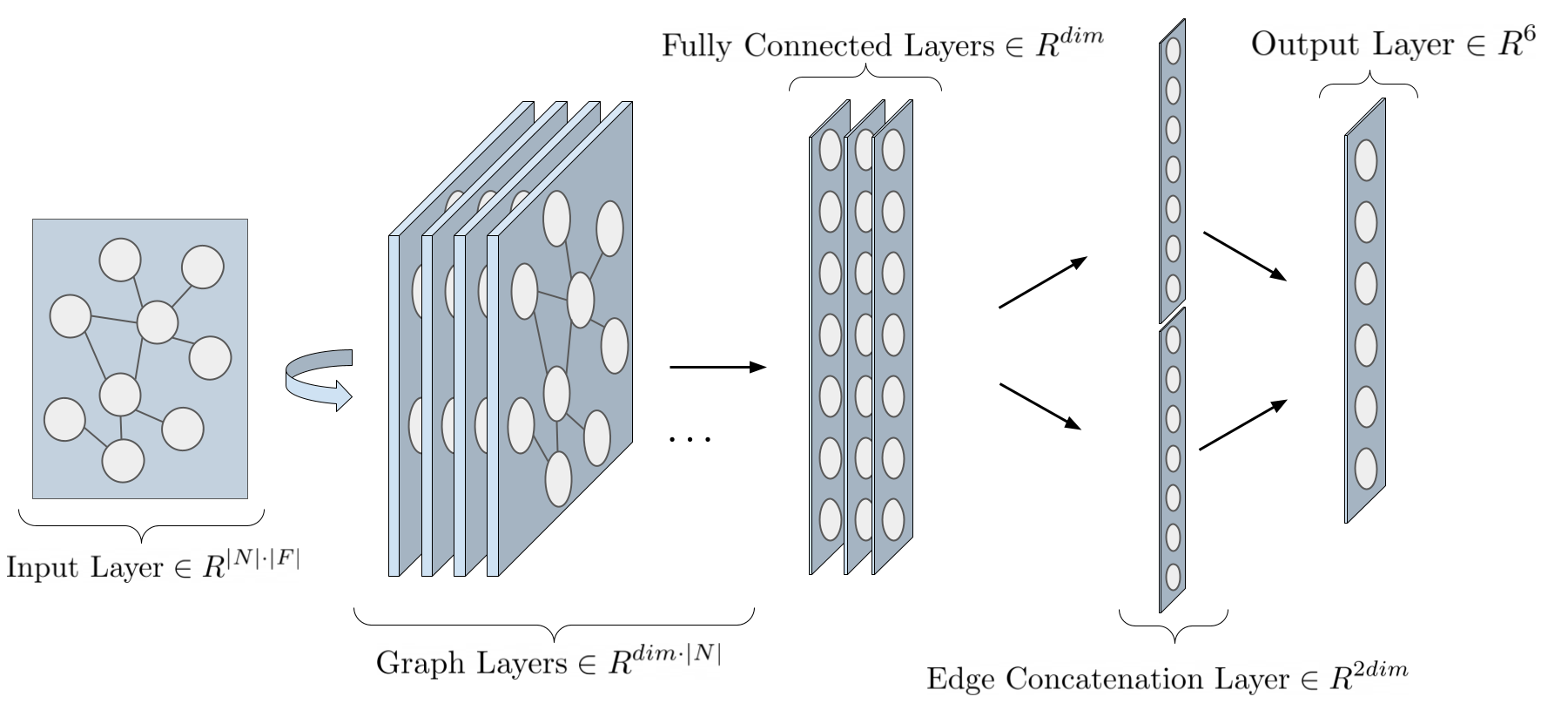}
         \caption{Overview of neural network architecture for graph neural networks. The number of graph layers (convolutional depth) and number of fully connected layers (linear depth) are hyperparameters. $|N|$ is the number of nodes in the input pathway. $|F|$ is the number of input features for each node.}
         \label{fig:nn_layout}
     \end{figure*}

We evaluated three general categories of models (Section~\ref{sec:methods_models}): general classifiers~\cite{scikit-learn}, probabilistic graphical models, and graph neural networks (Figure~\ref{fig:nn_layout}).
The fully-connected neural network (FullyConnectedNN), random forest (RF), and logisitic regression (Logit) served as baseline classifiers because they use no topological information from the pathway graph (Figure~\ref{fig:trad_classifiers}).
These models instead concatenate the node features of each interaction's endpoints as their input.
All other models use topological information from the pathway graph to encourage interactions near each other to have similar localizations.

%\underline{Fully connected neural network:} 
%In order to investigate the effects of graph topology on localization prediction, a fully connected neural network was constructed. 
%The linear neural network begins with an edge concatenation layer as shown in Figure \ref{fig:nn_layout}, followed by fully connected layers of size $dim$ up to \emph{linear depth}. 
%%%% Okay to keep these details in the supplement, and the reference to the figure was a bit confusing because only the edge concatenation part is relevant but the rest of the model architecture is different

\textbf{Graph convolutional network (GCN):}
Graph convolutional networks~\cite{gcn2017} incorporate a series of message-passing convolutional layers before the final fully connected layers. 
The convolutional layers allow for information to be shared across the topology of the input network, providing a first-order approximation of spectral graph convolutions~\cite{spectralApprox}. 
All neural network models were implemented using PyTorch Geometric~\cite{pytorchgeo}. 

\textbf{Graph attention network (GAT):} 
Graph attention networks extend graph convolutional networks by allowing each node to choose which neighbors to pay attention to. 
As opposed to taking the average of its neighbors, each node computes a weighted average of its neighbors in graph convolutional layers~\cite{velickovic2018gat, brody2021attentive}.  
The GAT is multi-headed, where multiple attention weights are computed in parallel for each node. 
The number of heads is a hyperparameter.

\textbf{Graph isomorphism network (GIN):}
Graph isomorphism networks~\cite{xu2018GIN} take advantage of the similarity between neighbor aggregation in graph neural networks and the Weisfeiler-Lehman (WL) graph isomorphism test~\cite{weisfeiler1968reduction}.
The WL graph isomorphism test is a heuristic algorithm for determining graph isomorphisms. 
The neighbor aggregation in each graph layer of a graph isomorphism network is formulated to be at least as powerful as the WL isomorphism test; the $l^{th}$ layer is guaranteed to generate different embeddings of two graphs if those graphs would be found to be non-isomorphic via the WL isomorphism test in $l$ iterations. 

\textbf{Probabilistic graphical models:}
Given the nature of the label propagation inherent in the pathway level localization prediction task, and that many localization databases provide scores or even probabilities, probabilistic graphical models are a natural choice.
However, these models only provide predictions on the nodes of the graph, while we are interested in localization labels on the edges. 
To convert the input pathway into an appropriate graphical model, each pathway is converted into a bipartite graph, where an additional node is added to that graph for each edge (Figure~\ref{fig:pgm_layout}).

Probabilistic graphical models represent a set of $N$ random variables $\mathbf{y}$ as nodes and dependencies between them as a set of edges $E$. 
We created two pairwise undirected probabilistic graphical models~\cite{UGMs}, which we call NaivePGM and TrainedPGM.
In these probabilistic graphical models the random variables obey a local Markov property, such that each random variable is conditionally independent of all others given its neighbors in the graph. 

The NaivePGM is a Markov random field, where protein localization database data is used to create conditional probability tables. 
In the TrainedPGM, input features are treated as observations of additional variables to train potential functions on each node. 
These potential functions are represented by discriminative classifiers~\cite{dgmlib}, here random forests.
This type of model is referred to as a discriminative random field~\cite{disRF_2006}.
This was chosen over a more traditional log linear parameterization due to better performance on the tuning data. 

We performed 30 iterations of hyperparameter selection via Bayesian optimization~\cite{bayesOpt_botorch} using Ax for neural network models and Scikit-optimize for classifier models\footnote{\url{https://ax.dev/} and \url{https://scikit-optimize.github.io/stable/}} (Tables \ref{tab:params} and \ref{tab:params_values}).
%Bayesian optimization was performed for $30$ iterations for each model.
%Fix Table S1 numbering (low priority but double check it is correctly hard-coded)

\section{Results}
\subsection{Comparing Pathway and Localization Databases}

To better understand the feasibility of predicting interaction localizations from protein-level localization data, we compared the edge localizations present in biological pathway databases to node localizations in protein localization databases.
The Reactome and PathBank pathway databases significantly disagree with both protein localization databases. 
For instance, among all proteins with an edge localized to the membrane in Reactome, ComPPI scores more as being in the cytosol than in the membrane.
In all cases there is a wide distribution when stratifying the ComPPI node scores used as features by the Reactome or PathBank edge localizations used as labels (Figures \ref{fig:comPPI_scoredist} and \ref{fig:compartments_scoredist}).
Therefore, for any individual protein and interaction there is a significant chance that protein's most likely localization according to ComPPI or Compartments is not the localization Reactome or PathBank assigned it to.

Directly using data from protein localization databases is not sufficient to accurately predict pathway level localization.
Many interactions have at least one contradictory interaction with an identical featurization but a different localization label, over $40\%$ when using ComPPI and over $20\%$ when using Compartments.  
In addition, many interaction localizations would be considered impossible when using a protein localization database alone. 
Almost $14\%$ of interactions in Reactome are between proteins that have no protein localizations in common in ComPPI. 
Even without featurization, for $9.5\%$ and $11.5\%$ of total interactions in Reactome and PathBank, respectively, there exists another interaction between the same unique proteins in another pathway that has a different localization. 
This indicates that pathway topology or some other form of additional information beyond that of individual proteins is needed to correctly predict localization in context.

\subsection{Pathway Database Localization Prediction}
\begin{figure*}[htbp]
     \centering
     \makebox[\textwidth][c]{\includegraphics[width=0.9\textwidth]{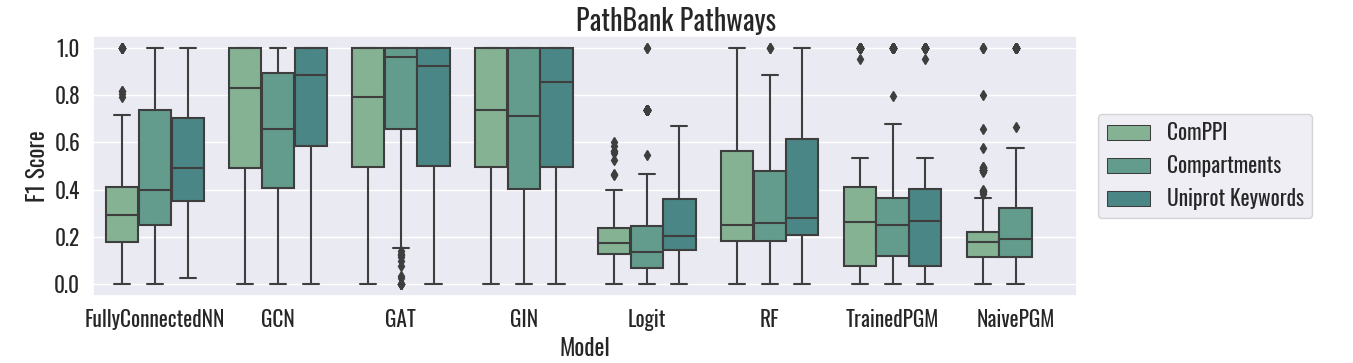}}
     \caption{Multiclass F1 score of predictive performance on PathBank localizations across all $427$ considered PathBank pathways. Scores are calculated per pathway; the distribution of scores is shown for each model.}
     \label{fig:pb_pathways_f1}
 \end{figure*}

\begin{figure*}[htbp]
     \centering
     \makebox[\textwidth][c]{\includegraphics[width=0.9\textwidth]{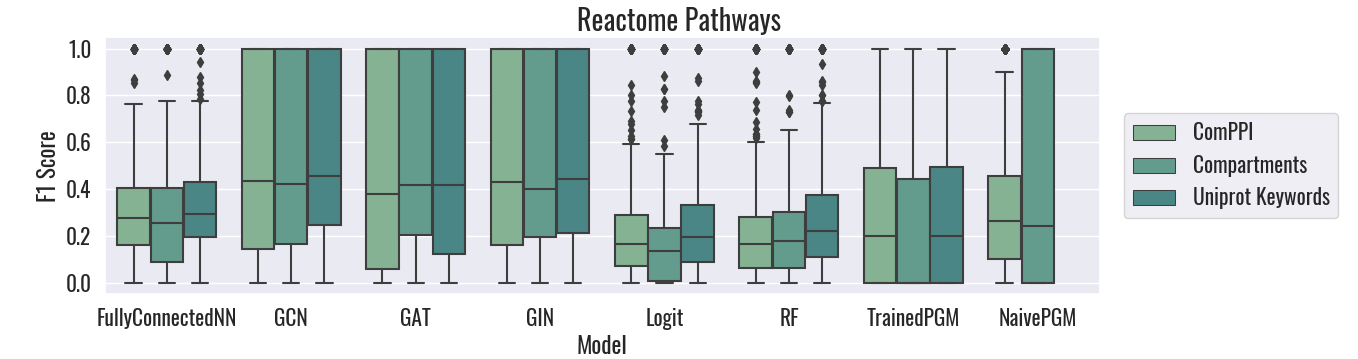}}
     \caption{Multiclass F1 score of predictive performance on Reactome localizations across all $918$ considered Reactome pathways. Scores are calculated per pathway; the distribution of scores is shown for each model.}
     \label{fig:reactome_pathways_f1}
 \end{figure*}

We used cross-validation to train our models on protein  information and some labeled database pathways and evaluate their edge localization predictions for other database pathways given only protein information and graph structure as input.
%Figure \ref{fig:pb_pathways_f1} and \ref{fig:reactome_pathways_f1} show performance in predicting interaction localizations in PathBank and Reactome pathways, respectively. 
Overall, models were able to achieve better interaction localization prediction performance on PathBank pathways (Figure~\ref{fig:pb_pathways_f1}) than Reactome pathways (Figure~\ref{fig:reactome_pathways_f1}).
Generally, models' performance in predicting PathBank interaction localizations was more consistent across pathways. 
However, on both datasets all models' performance had high variance across pathways.
Except for logistic regression, all models got at least some pathways completely correct and some pathways completely wrong across all databases and feature sets. 
The graph neural network models, GCN, GAT, and GIN, generally outperformed other models in all conditions.
However, in Reactome no model was able to achieve a median multiclass F1 score (hereafter called `F1 score') of over $0.5$
%After submission, it would be nice to consider some type of randomized control where the graph methods are run on shuffled version of the pathway edges. Is it the actual pathway structure that improves F1 score or simply the ability to smooth among a set of connected nodes that would remain connected even if edges are randomized?

Probabilistic graphical models and models that used no pathway topology had generally comparable performance. 
The FullyConnectedNN model was able to outperform other models when predicting PathBank localizations using Compartments or UniProt keyword features. 
It should be noted, however, that when calculating performance by pathway as done in this setting, the size of each pathway is not taken into account.
This means that edges in very small pathways can have an outsized effect on total performance. 

Alternatively, Figures \ref{fig:merged_pb_f1} and \ref{fig:merged_reactome_f1} show F1 scores for each model aggregated from all pathways, where all edges are used for a single performance calculation.
When aggregated in this way, all non-neural network models perform comparably.
The probabilistic graphical models, and the TrainedPGM model in particular, struggled with small pathways. 

The number of real and predicted unique localizations in each pathway also had a large effect on model performance. 
This can be thought of as the smoothness of the real or predicted localizations in a pathway, or how strong the tendency is for edges nearby in a pathway to have the same localization. 
Ideally, a model would be able to detect that a pathway exists entirely in a single localization and aggressively smooth its localization predictions over the pathway. 
Pathways with a single localization had the widest range of performance within each model. 
More extreme performances, at or nearly at 1.0 or 0.0 for these pathways, indicate that the model correctly predicted that the pathway had only a single localization. 
Figure \ref{fig:unique_dists} shows the distributions of the number of predicted unique localizations by the different models.

\subsection{HCMV Infection Spatial Proteomics Case Study}   

We considered three scenarios for evaluating localization prediction in an experimental setting. 
Here, we examine if localizations can be inferred in the context of a HCMV infection (Section~\ref{sec:methods_casestudy}). 
We simulate an exploratory workflow by first constructing HCMV infection-specific biological pathways using pathway reconstruction~\cite{omicsIntegrator} (example pathway topologies can be viewed in Figures \ref{fig:best_120_pathway} and \ref{fig:best_egf_pathway}). 
We then use the context provided by these pathways' topologies to predict interaction localizations with the best performing model from pathway database prediction, GAT, using node features from the Compartments database.

In all scenarios, we predict localizations for each interaction of pathways created from protein abundance measurements at 120hpi. 
Localization data from spatial MS taken at the same timepoint was used as ground truth.
Each scenario differs in the labeled training data used: pathways from a pathway database, a different experiment using a different context and cell type, or data from the same experiment at a different timepoint. 
In all scenarios, all data from the 120hpi timepoint was held out until the final evaluation. 
We also consider a baseline model that always predicts the most frequent localization among all training set interactions. 

\begin{figure*}[htbp]
         \centering
         \includegraphics[width=0.7\textwidth]{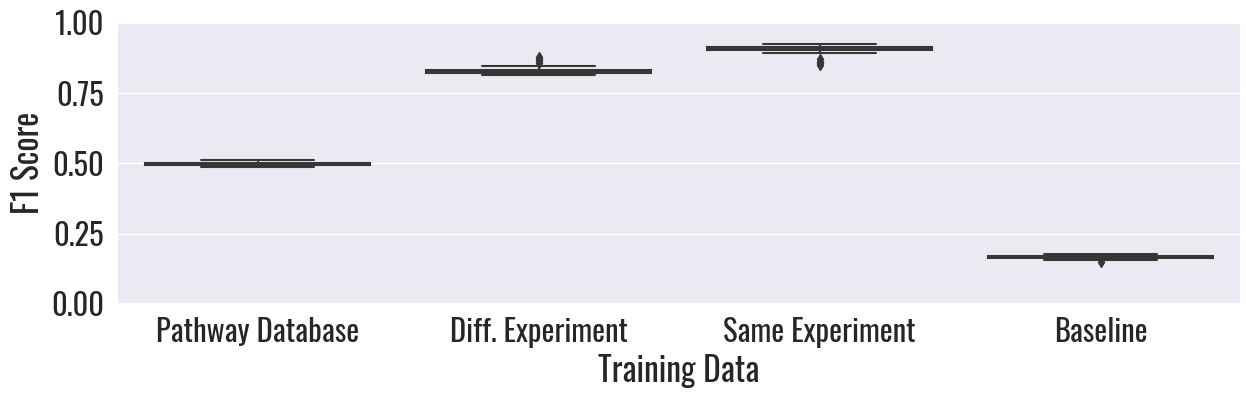}
         \caption{Multiclass F1 score of the GAT model on spatial MS data of viral infection at 120hpi. Performance is shown in each scenario for the $50$ top pathways created from a parameter sweep. The baseline model always predicts the most common localization in the training dataset. }
         \label{fig:caseStudyPerf}
     \end{figure*}

While in all scenarios the model substantially outperformed the baseline, there was a large gap in performance between the model trained using pathway databases versus those trained on a different experiment (Figure \ref{fig:caseStudyPerf}). 
Both scenarios using experimental data achieved an F1 score of over $0.8$.
Although the GAT model predictions do not perfectly recapitulate the spatial proteomics localizations, it is encouraging that the GAT model trained in a plausible setting with data from an unrelated biological context is almost as accurate as the unrealistic, best case GAT model trained on another timepoint from the same HCMV infection experiment.
%The highest performing model used data from the same experiment. 
%However, it should be noted that this scenario is not realistic in a typical biological discovery workflow. 
    
\section{Conclusions and Future Work}

Although there is some correspondence between protein localization databases and localization data in pathway databases, these two types of localization data generally disagree. 
Graph neural network models were required to achieve high predictive performance on PathBank localizations, and all models performed poorly in predicting Reactome localizations.

There are a number of possible reasons for this misalignment between localization information in pathway databases and protein localization databases.
While the best-performing models include topological information, implying that topology is needed to bring context to protein localization, it is possible that other types of data are needed. 
Protein features derived from UniProt keywords only slightly improved performance, and tissue- or cell-specific localization may be necessary to fully realize context-specific localization.
That type of information may not be available for pathway databases, which are often provided independent of tissue type, but could be for reconstructed pathways.
%However the possible necessity of contextual information such as tissue type, when most pathways in pathway databases are provided independent of tissue type, may be infeasible.  
The protein localization databases may also be too noisy and general for context-specific localization prediction. 
While some signal does exist, the wide range of distributions for ComPPI and Compartments scores across different pathway localizations highlights the imprecise nature of the prediction problem. 
%It may be the case that the predictive task needs features that also include contextual information.

While graph neural networks outperformed other methods in predicting pathway localizations, it is unclear how large a role pathway topology played in these methods' performance. 
It is possible that increased performance over other models comes solely from how graph convolutions share information between nodes, as opposed to the biological information inherent in each pathway's topology aiding localization prediction.
% This text is a good short summary. If space permits, we could note why randomizations are hard in this setting because we make predictions on the edges. We could also point out the PGMs also share information but don't do as well.

%PathBank is only partially labeled with subcellular localizations, while Reactome is fully labeled, which may explain part of the difference in performance on the two databases.
%While neither database makes their reasoning explicitly clear for their strategy to including subcellular localizations, it may the the case that, as Reactome requires all interactions to have a localization, 
%We speculate that Reactome's pathways may be forced to include localization information even when an interaction's subcellular location may not be known with high confidence. 
%PathBank, however, can choose to omit low-confidence or unknown interaction localizations.

The conversion of pathways from hypergraphs to graphs greatly impacted the class distribution and topology of Reactome and PathBank pathways.
Treatment of protein complexes can lead to orders of magnitude difference in the number of edges in the resultant pathways. 
%Consideration of the desired analysis task is important when making this decision.
%For instance, the conversion we chose, creating protein complex nodes to represent complexes, removes node information but better preserves the edge structure and balance in the pathway. 
We created protein complex nodes to represent complexes, which removes node information but better preserves the edge structure and balance in the pathway.
An analysis task focused specifically on nodes may want a conversion that better preserves node information at the possible cost of edge information.
Important future work would be to consider these conversions in a more systemic way and quantify the hypergraph properties they alter or keep invariant. 

% An alternate direction in the treatment of hypergraphs would be moving prediction directly onto the hypergraph, removing any conversion step.
% While hypergraph neural networks exist~\cite{hypergraphs_gao2019,hypergraphs_yadati2019}, they are much less mature than graph neural networks.
% However, this would preserve information lost from the hypergraph conversion.

Pathway reconstruction has already proven to be a powerful strategy for interpreting transcriptomic, proteomic, or other data in a network context, and the ability to coarsely approximate interaction localizations could further increase its value.
We observed the GAT model may have sufficient accuracy to roughly estimate such pathway localizations as long as it is trained on experimental data instead of pathway databases.
Predictions using the model trained on HeLa cells still had an error rate of approximately $17\%$ but could plausibly be used to obtain an estimate of context-specific localization predictions in the absence of other data. 
Further testing is required to assess how similar the training conditions and assay types must be to the test conditions and assays and what types of pathway reconstruction algorithms are compatible with our GAT localization prediction model.

%The GAT model was able to achieve high performance in directly predicting experimental data, if also trained using experimental data.
%While it is unrealistic in an actual experimental scenario to have data from an identical cell type, condition, and laboratory, the scenario using experimental data from HeLa cells shows a plausible use of graph neural networks for predicting context-specific localization.
%This is a promising direction for being able to add localization information to exploratory analyses using reconstructed biological pathways.

There are additional biological contexts where localization prediction could prove valuable. 
Single-cell spatial proteomics experiments have previously found proteins to vary by as much as $16\%$ in either expression or spatial distribution between cells undergoing the same process in the same tissues~\cite{thul_subcellular_2017}. 
Predicted protein localizations for individual cells could add an additional layer of information in single-cell analyses. 
Additionally, targeted identification of abnormal protein localizations could provide insight in diseases where protein localization is known to play a role~\cite{bliseSingleCellSpatialProteomics2021}. 
The current predictive method could be expanded to attempt to quantify a localization being unexpected given a constructed pathway representing some cellular state.

% May want to remove `eprint' so the long URLs aren't included
% Problem with beta character in D’Souza reference
\bibliographystyle{unsrt2}

\bibliography{main}

\section{Acknowledgements}
This work was supported by NIH award T15LM007359, NSF award DBI 1553206, the Morgridge Institute for Research, and the University of Wisconsin–Madison Office of the Vice Chancellor for Research and Graduate Education with funding from the Wisconsin Alumni Research Foundation. We thank Sushmita Roy for her valuable feedback.

\clearpage

\pagestyle{empty}

\begin{center}

\huge Supplementary Information: Graph algorithms for predicting subcellular localization at the pathway level

\vspace{5mm}

\large  Chris S Magnano and Anthony Gitter 

\end{center}

\setcounter{page}{1}
% Prefix section, table, and figure numbers with S in the supplement
\setcounter{table}{0}
\renewcommand{\thetable}{S\arabic{table}}
\setcounter{figure}{0}
\renewcommand{\thefigure}{S\arabic{figure}}
\setcounter{section}{0}
\renewcommand{\thesection}{S\arabic{section}}

\section{Supplementary Methods}

%\subsection{Experimental Setup}
\subsection{Spatial Proteomics Case Study}
\label{sec:methods_casestudy}

\paragraph{Mass spectromety datasets:} To demonstrate a use case of pathway-based localization prediction, and validate its performance, we performed localization prediction on data from a study that measured proteomic quantification of primary fibroblasts during human cytomegalovirus (HCMV) infection.
Two mass spectromety quantification methods were used at five timepoints: 24, 48, 72, 96, and 120 hours post infection (hpi). 
The first of these datasets provides label-free protein quantification at each timepoint. 
The other was quantified using isobaric labeling via tandem mass tags (TMTs). 
These two datasets will be referred to as the label-free and the TMT datasets, respectively.

Multi-organelle profiling was performed on the TMT dataset via gradient centrifugation to fractionate organelles.
This process partially separates organelles into a set of subcellular fractions.
While each of these fractions is not purely a single organelle, each organelle contains a unique signature in its quantification across the subcellular fractions. 
In the original analysis, clustering analysis was then used to group proteins with similar fraction profiles.
Finally, each cluster was labeled as a certain organelle via a set of marker proteins, proteins known with high-confidence to localize to a particular organelle.

In the HCMV protein quantification, a set of marker proteins was curated from UniProt subcellular location annotations with experimental evidence; proteins that were annotated with multiple localizations were excluded from the marker set. 
Proteins that were not confidently assigned to a particular organelle were left as unlabeled in the original study's localization labeling. 
From this, there were 2,730 proteins in total, of which 1,229 had localization labels at 24hpi and 1,348 had labels at 120hpi. 
$574$ of these proteins were marker proteins. 

\paragraph{Experimental scenarios:}
We explored three different data availability scenarios for using the best performing model from the pathway database prediction task, GAT, to predict localizations in an experimental setting. 
For all three scenarios, the best performing hyperparameter combination from the pathway database prediction task was used with no further hyperparameter tuning. 
The three scenarios are as follows:

\begin{enumerate}
    \item First, we trained a model using data from the PathBank database as described in Section \ref{sec:methods_locPred}. Here we took the best performing pre-trained model with no further modifications and used it to predict localizations at the 120hpi timepoint. 
    \item Second, we trained a model using a separate dataset that measured protein localization using a similar method on a different cell type and under a different biological condition, HeLa cells undergoing EGF stimulation.
    \item Third, we trained a model on the same HCMV infection experiment at the 24hpi timepoint. 
    This third scenario is impractical, as it would require a dataset to already exist for an identical cell type and condition. However, it  gives a useful benchmark for predictive performance versus the practical second scenario. 
\end{enumerate}

\begin{figure*}[htbp]
         \centering
         \makebox[\textwidth][c]{\includegraphics[width=\textwidth]{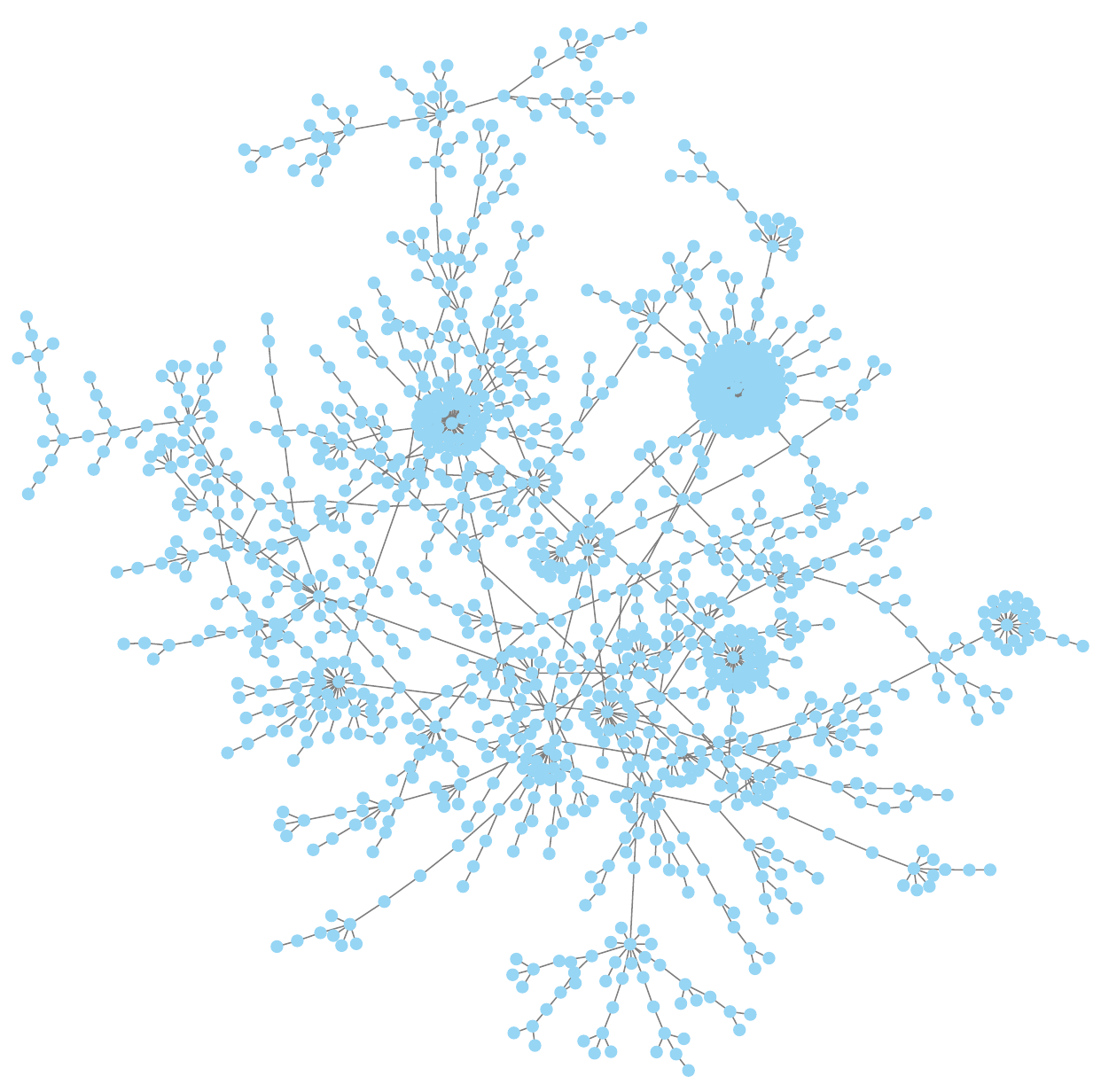}}
         \caption{Topology of the best ranked pathway reconstruction of HCMV infection at 120hpi, containing 1,226 interactions. Pathways were reconstructed using Omics Integrator 2.}
         \label{fig:best_120_pathway}
     \end{figure*}

\begin{figure*}[htbp]
         \centering
         \makebox[\textwidth][c]{\includegraphics[width=0.75\textwidth]{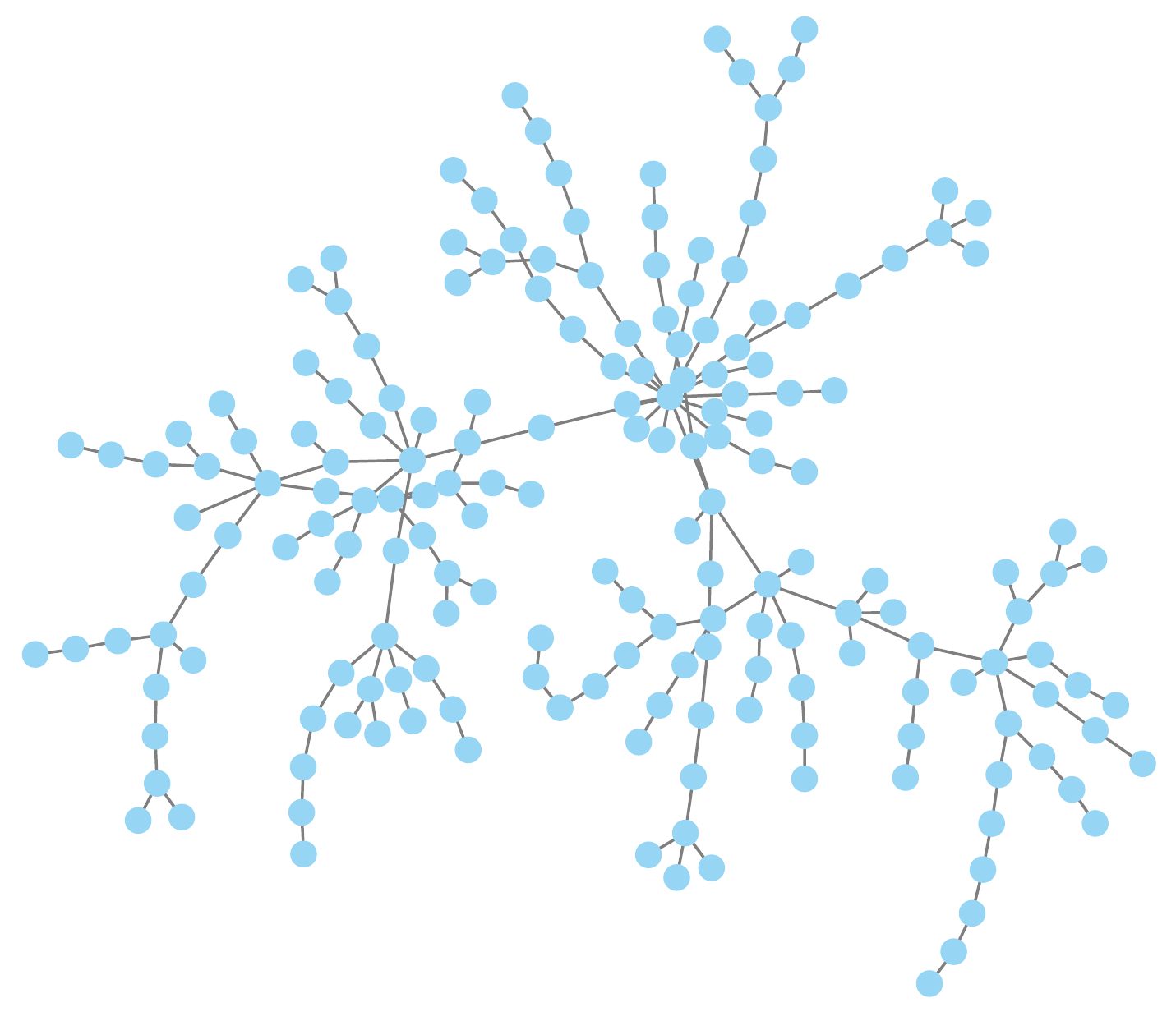}}
         \caption{Topology of the best ranked pathway reconstruction of EGF stimulation, containing 187 interactions. Pathways were reconstructed using Omics Integrator 2.}
         \label{fig:best_egf_pathway}
     \end{figure*}

\paragraph{Pathway reconstruction:} Pathway reconstruction was performed using Omics Integrator~2\footnote{\url{https://github.com/fraenkel-lab/OmicsIntegrator2}}, which is similar to Omics Integrator but uses a different optimization algorithm\footnote{\url{https://github.com/fraenkel-lab/pcst_fast}}.
Omics Integrator 2 performs pathway reconstruction via the prize-collecting Steiner forest problem.
Omics Integrator 2 was run using the Signaling Pathway Reconstruction Analysis Streamliner (SPRAS)\footnote{\url{https://github.com/Reed-CompBio/spras}}.
Protein abundance fold-change was used as prizes for both fibroblast HCMV infection data and HeLa EGF stimulation data.
The background protein interaction network for pathway reconstruction was a previously published network with 161,901 weighted edges based on merged interactions from the iRefIndex database v13\footnote{\url{https://irefindex.vib.be/wiki/index.php/iRefIndex}} and kinase-substrate interactions from PhosphoSitePlus\footnote{\url{https://www.phosphosite.org}}.
The Omics Integrator 2 hyperparameter $\omega$ was tested between $1$ and $10$, $\beta$ was tested between $1$ and $5$, and $\mu$ was tested between $0.1$ and $1$. 
Each hyperparameter was evaluated at $10$ increments across its range.
For the 120hpi timepoint, pathway parameter advising was used to select the top 50 pathways from 1,000 candidate hyperparameter combinations.
For the 24hpi timepoint and EGF stimulation datasets, all nonempty pathways were used so that these pathway datasets were comparable in size to the pathway database datasets, resulting in training sets of 542 and 503 pathways, respectively. 
Reconstructed pathways from HCMV infection were large, with an average of 1,279 interactions per pathway at 120hpi.
Other pathway sizes were more similar, with EGF stimulation pathways, Reactome pathways, and PathBank pathways having an average of 202, 91, and 229 average interactions, respectively. 
The top ranked pathway at 120hpi and for EGF stimulation can be viewed in Figures \ref{fig:best_120_pathway} and \ref{fig:best_egf_pathway}, respectively.
All GAT models' hyperparameters were the same as those used in the highest performing GAT model when using Compartments features to predict PathBank labels, as recorded in Table \ref{tab:params_values}. 

\subsection{Pathway Localization Prediction Models}

\label{sec:methods_models}

\subsubsection{Neural Networks}

The maximum value in the output layer was used as the final class label prediction. 
All neural network models were trained using cross-entropy loss. 
All neural networks were implemented using PyTorch and the PyTorch Geometric package.

\paragraph{Graph convolutional network:}
The graph convolutional network incorporated a set of message-passing convolutional layers before the final set of fully connected layers. 
These convolutional layers allow for information to be shared across the topology of the input network. 
The $l^{th}$ convolutional layer $H^{(l)}$ is updated via the following rule:

$$H^{(l)} = ReLU(D^{-\frac{1}{2}}\widetilde{A}D^{\frac{1}{2}}H^{(l-1)}W^{(l-1)})$$

Where $\widetilde{A}$ is the adjacency matrix of the input pathway with added self-edges for all nodes, $D$ is a degree matrix normalization factor where $D_{ii} = \sum\limits_{j} \widetilde{A}_{ij}$, and $W^{(l)}$ is a set of weights for the $l^{th}$ layer.  
This update rule provides a first-order approximation of spectral graph convolutions and is implemented in the $GCNConv$ class in PyTorch Geometric.

\paragraph{Graph attention network:} 
Graph attention networks extend graph convolutional networks by allowing each node to choose which of its neighbors to pay attention to. 
As opposed to taking the average of its neighbors, each node computes a weighted average of its neighbors in graph convolutional layers.  
Furthermore, many attention networks are multi-headed, where multiple attention weights are computed in parallel for each node. 
The number of heads to include is an input hyperparameter and often increases accuracy at the cost of increased computational complexity.
We used the PyTorch Geometric $GATV2Conv$ class for graph layers, which is a more expressive implementation of graph attention that allows for more diversity in attention between nodes. 

\paragraph{Graph isomorphism network:}
Graph isomorphism networks take advantage of the similarity between neighbor aggregation in graph neural networks and the Weisfeiler-Lehman (WL) graph isomorphism test.
The WL graph isomorphism test is a heuristic algorithm for determining graph isomorphisms. 
For two graphs, in each iteration of the test every node aggregates its neighbors into a unique hash. 
These hashes are compared between the two graphs, and if they differ the graphs are known to be non-isomorphic. 
Iterations of the test are repeated until the user feels confident that the graphs are isomorphic; the algorithm cannot conclusively prove isomorphism. 

The neighbor aggregation in each graph layer of a graph isomorphism network is formulated to be at least as powerful as the WL isomorphism test; the $l^{th}$ layer is guaranteed to generate different embeddings of two graphs if those graphs would be found to be non-isomorphic via the WL isomorphism test in $l$ iterations. 
The representation of each node $n$ in layer $l$ of a graph isomorphism network, $h_n^{(l)}$, is computed as:

$$h_n^{(l)} = MLP^{(l)} \left((1 + \epsilon^{(l)}) \cdot h_n^{(l-1)} + \sum\limits_{u \in Adj(n)} h_{u}^{(l-1)}\right)$$

Where $MLP$ is a multi-layer perceptron, $\epsilon$ is a learned parameter, and $Adj(n)$ is the set of nodes adjacent to $n$ in the input pathway. 
We used the $GINConv$ class in PyTorch Geometric for graph isomorphism layers. 

\subsubsection{Probabilistic Graphical Models}
% Double check tense consistency
Given the nature of the label propagation inherent in the pathway level localization prediction task, and that many localization databases contain uncertain or even probabilistic data, probabilistic graphical models are a natural choice. 
As moving between subcellular locations costs energy, it is unlikely to happen often within a single pathway. 
Therefore, we can make the assumption from a modeling perspective that the subcellular location of an interaction is dependent on the subcellular location of neighboring interactions within a pathway.

Probabilistic graphical models represent a set of $N$ random variables $\mathbf{y}$ as nodes and dependencies between them as a set of edges $E$. 
We created two types of pairwise undirected probabilistic graphical models, which we call NaivePGM and TrainedPGM.
In these probabilistic graphical models, the random variables obey a local Markov property, such that each random variable is conditionally independent of all others given its neighbors in the graph. 

The NaivePGM is a Markov random field, where the joint probability of all localizations can be factorized as 

$$P(\mathbf{y}) = \frac{1}{Z} \prod\limits_{i \in N} \phi_{i}(y_i) \prod\limits_{i,j \in E} \phi_{ij}(y_i, y_j)$$

Where $Z$ is a normalizing function so that the combination of all possible configurations of $\mathbf{y}$ sum to 1 and $\phi_{i}({y_i})$ and $\phi_{ij}({y_i,y_j})$ are the unary and pairwise potential functions, respectively. 
The unary potential function defines the probabilities of a given node having each localization, while the pairwise potential functions define the joint probability of each pair of nodes that share an edge.
For finding the task of finding the most likely configuration of $\mathbf{y}$, referred to as decoding, $Z$ can be ignored.
In the NaivePGM, the input features are not used to parameterize the potential functions. 
Instead, the unary potential functions directly map the normalized features to class probabilities, and the joint probability tables directly map the normalized features to joint probability tables. 
This was chosen because both the ComPPI and Compartments scores represent confidences, with ComPPI scores directly representing probabilities for each localization.
However, this use of scores means that the NaivePGM cannot use Uniprot keyword-derived features as they do not represent localization confidence.  

The TrainedPGM is a conditional random field where the input features are treated as observations of additional variables. The probability of localization assignments $\mathbf{y}$ are then conditioned over the input features $\mathbf{x}$ as:

$$P(\mathbf{y}|\mathbf{x}) = \frac{1}{Z} \prod\limits_{i \in N} \phi(x_i, y_i) \prod\limits_{i,j \in E} \phi(x_i, y_i, y_j)$$

Here, the unary potential functions are now conditioned on observations of features $x_i$ corresponding to each random variable $y_i$. 
Edge potentials represent the dependence between each node's state $y_i$ and its neighbor's state $y_j$ given its features $x_i$.
Each random variable $y_i$ is conditionally independent of all other variables given its corresponding features $x_i$ and its neighbors' states $y_j$. 

The potential functions in conditional random fields are typically log-linear functions of the form $e^{w_i^{T}\phi_{f}(x_{i},y_{i})}$, parameterized via a weight vector $\mathbf{w}$, and $\phi_f$ simply represents features for each node. 
Additionally, typically the feature weight vectors are shared between nodes or sets of nodes.
Thus, the entire model can then be represented as:

$$p(\mathbf{y}|\mathbf{x},\mathbf{w}) = \frac{1}{Z} exp\bigg(\sum\limits_{i \in N} \mathbf{w}_{f}^{T} \phi_{f}(x_i,y_i) + \sum\limits_{i,j \in E} \mathbf{w}_{e}^{T} \phi_e(x_{i}, y_i, y_j) \bigg)$$

Where $\phi_{f}(x_{i},y_{i})$ is the single unary potential function that represents features for each node, and $\phi_{e}(x_{i},y_{i},y_{j})$  is the single pairwise potential function that represents combinations of states.
The weight vector $\mathbf{w_{f}}$ is a set of weights for each feature to each possible configuration of $y_i$, while the weight vector $\mathbf{\mathbf{w}_{e}}$ is a set of weights for each feature to each combination of configurations for $y_i$ and $y_j$. 

When represented with these potentials, the log likelihood of the model parameters $\mathbf{w}$ can be easily represented, and is differentiable, allowing for parameters to be learned by maximum likelihood estimation via gradient-based optimization. 
Sets of nodes and edges can share the same set of model parameters, referred to as parameter tying. 
However, parameter learning for a conditional random field of this form did not converge when trained with stochastic gradient descent. 
This may be due to the underlying label distributions of different pathways being too different from each other. 

Instead an alternative model formulation was chosen where potentials are represented by discriminative classifiers, specifically random forests.
This type of model is referred to as a discriminative random field.
This was chosen over a more traditional log linear parameterization due to better performance on tuning data during model selection. 

\begin{figure*}[htbp]
          \centering
          \includegraphics[width=\textwidth]{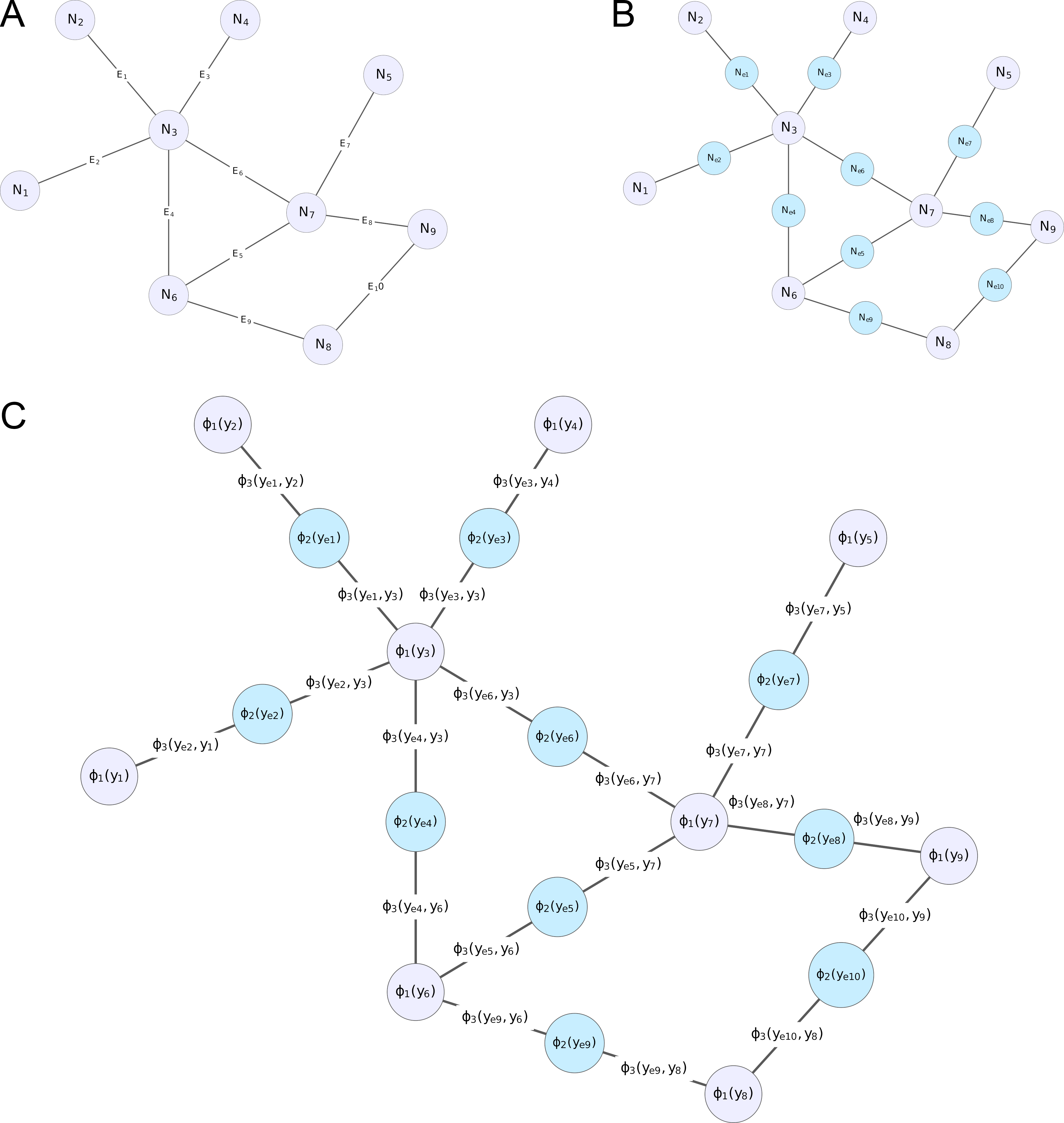}
          \caption{Overview of how pathways were represented as probabilistic graphical models for interaction classification. 
          Panel A shows the original pathway structure. Panel B shows the variables that are added to the graphical model to represent interactions in the pathway.
          Finally, Panel C shows how potential functions are used and tied. There are 2 sets of unary potentials, $\phi_1()$ and $\phi_2()$, which model the original nodes and the interaction nodes, respectively. $\phi_3()$ models how each interaction relates to its adjacent nodes. }
          \label{fig:pgm_layout}
      \end{figure*}

A traditional construction of a probabilistic graphical models from the nodes and edges of a biological pathway would only provide predictions on the nodes of the graph, while we are interested in localization labels on the edges. 
To convert the input pathway into an appropriate graphical model, each pathway is converted into a bipartite graph, where a node is added that represents each pathway edge. 
First, all nodes from the original pathway are added to the graphical model. 
No edges from the original pathway are added. 
Instead, for each pathway edge $e_{ij}$ between pathway nodes $n_i$ and $n_j$, a graphical model node $n_{eij}$ is added representing the interaction. 
Then two edges are added to the graphical model going from each original node to the new interaction node, $n_{ie_{ij}}$ and $n_{je_{ij}}$.
An overview of this process can be found in Figure~\ref{fig:pgm_layout}.
Each of the $K$ features $f_{ek}$ in $n_{eij}$ are computed as the normalized product of features from $n_i$ and $n_j$, here represented as $f_{ik}$ and $f_{jk}$:

$$f_{ek} = \frac{f_{ik}f_{jk}}{\sum_{l=1}^{K}f_{il}f_{jl}}$$

This is equivalent to how interaction localization probabilities are calculated in ComPPI.
Parameters are tied such that all original nodes are represented by one set of model parameters, and all interaction nodes are represented by another. 
This can be seen in panel C of Figure~\ref{fig:pgm_layout}, where $\phi_1$ is the set of model parameters that describes the relationship between input features for each protein and its localization, $\phi_2$ describes the relationship between each interaction's combined features and its localization, and $\phi_3$ describes the relationship between each protein and the interactions it participates in. 

Final localization labels can be viewed as a maximum a posteriori (MAP) estimate of the configuration of all interaction node labels. 
Decoding was performed using loopy belief propagation, which approximates the MAP estimate via a message passing algorithm.
Loopy belief propagation was run for 10,000 iterations in all cases. 
Both the NaivePGM and TrainedPGM models were implemented in the Direct Graphical Models software library\footnote{\url{https://research.project-10.de/dgmdoc/index.html}} v1.7.0. 

\subsubsection{Other Classification Models}

\begin{figure*}[hbtp]
         \centering
         \includegraphics[width=0.5\textwidth]{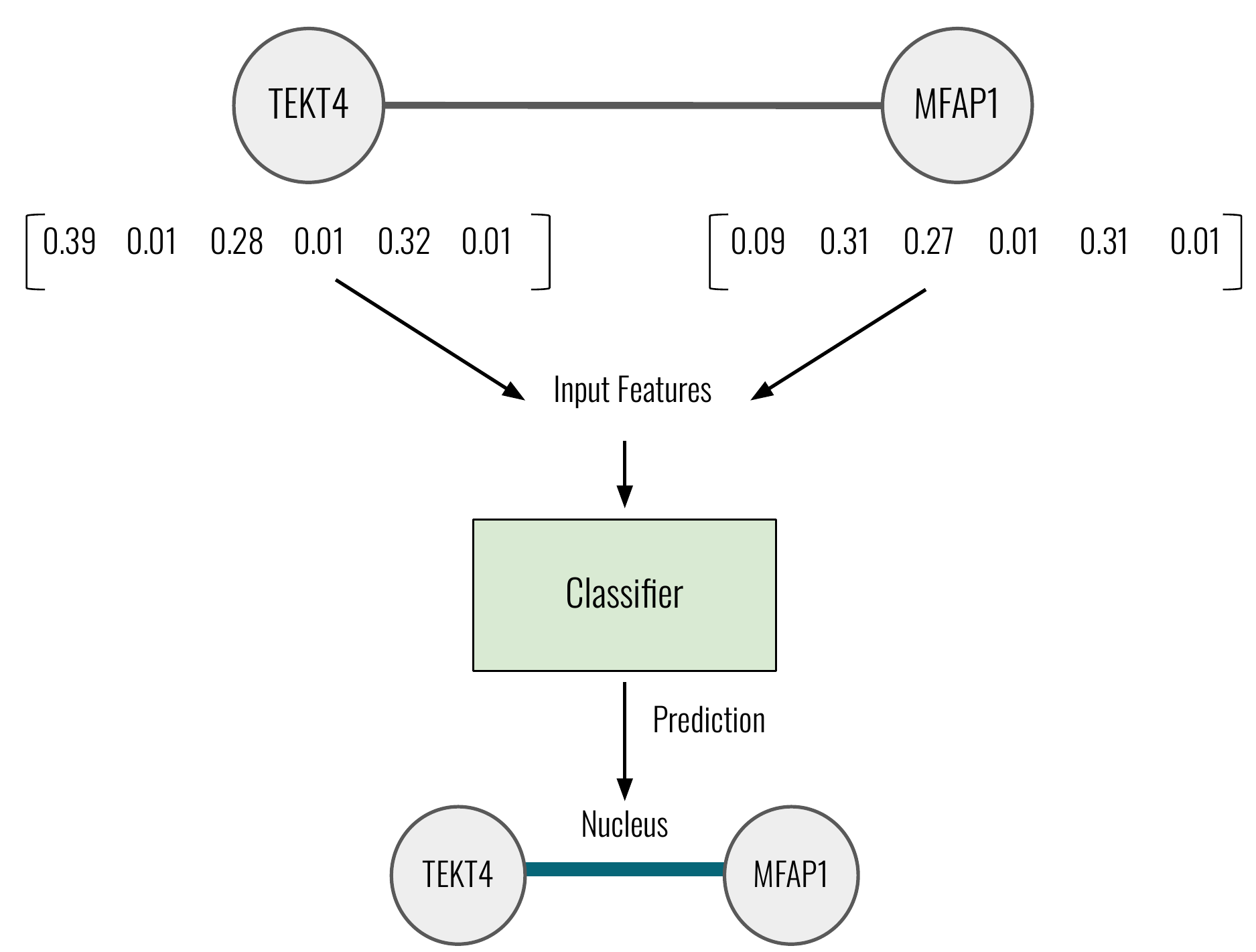}
         \caption{Overview of how topology-free classifiers are used for the edge labeling task of localization prediction.}
         \label{fig:trad_classifiers}
     \end{figure*}

Two non-neural network classifiers were used to further examine the effect of incorporating topological information into localization prediction: logistic regression, referred to as Logit, and random forests, referred to as RF. 
Interactions were represented by concatenating the features of the two nodes that make up that interaction in alphanumeric order by protein identifier, as shown in Figure \ref{fig:trad_classifiers}. 
Tested hyperparameter ranges for these models are listed in Table \ref{tab:params}.
Both models were implemented in Python 3.9 using the Scikit-Learn package v1.0.2. 

\subsection{Data}
\subsubsection{Pathway Databases}
\label{sec:pw_datbases}
Pathway datasets were constructed from the Reactome and PathBank databases. 
Pathways were downloaded from Pathway Commons, and localization information was retrieved from BioPax pathway representations using the PyBioPax package v0.1.0\footnote{\url{https://github.com/indralab/pybiopax}}.
Reactome contains localization information for all edges. 
In PathBank, however, approximately 9\% of edges have missing localization information. 
These edges with missing data were excluded from all analyses.

Localizations in Reactome and PathBank are given as cellular component Gene Ontology (GO) terms.
We mapped these GO terms to one of the six localizations in ComPPI by a combination of ComPPI's manually curated GO term mapping\footnote{\url{https://comppi.linkgroup.hu/help/subcell_locs}} and the GO term hierarchy. 
If a GO term had no mapping, we used its parent GO term's localization mapping.

After the protein-complex collapsing step (Section \ref{sec:exp_path_data}), all pathways with fewer than 4 nodes were excluded from the analysis. 
This resulted in 953 Reactome pathways and 467 PathBank pathways.

Both pathway databases contain a highly skewed distribution of localizations across all interactions. 
The rarest localization in both databases, secretory-pathway and nucleus in Reactome and PathBank, respectively, occurs in less than $0.5\%$ of all edges. 
The most common localization, which is cytosol for both databases, consists of $38\%$ of Reactome interactions and $52\%$ of PathBank interactions. 

\subsubsection{Protein Localization Databases}
\label{sec:loc_datbases}
%Two protein localization databases were used throughout all experiments, Compartments and ComPPI.

ComPPI is a meta-database for protein subcellular localizations.
It combines data from 8 subcellular localization databases. 
It does not include data from Compartments. 
Proteins are assigned scores for each of 6 subcellular locations: cytosol, plasma membrane, mitochondrion, extracellular, nucleus, and secretory-pathway. 
These 6 locations were used for all predictions; localizations for all other data sources were mapped to these 6.
ComPPI combines weights for different types of evidence across its data sources to give the probability of a protein to be found in a particular subcellular location. 
All human ComPPI data was retrieved on 2020-11-09. 

Compartments is a protein subcellular localization database that combines data from 4 different types of data: database annotations, experimental screens, automated text mining, and predictive sequence-based models. 
Each data source is given a confidence score between 1 and 5 based on the level of evidence. 
Compartments assigns proteins to 1 of 11 subcellular locations. 
These 11 locations were mapped to the 6 localizations in the ComPPI database. % Were all 11 able to be mapped or were some removed?
All Compartments data was retrieved on 2021-09-29.

\subsubsection{UniProt Keyword Features} 
\label{sec:uniprot_kw_feats}

To explore the value of additional protein data in predicting localization data, UniProt keywords were collected for all human proteins. 
UniProt keywords are a controlled hierarchical vocabulary that represent a variety of protein categories such as molecular function, disease participation, structural features, and biological processes. 
These keywords are manually assigned and include localization data. 
While UniProt keywords provide a range of protein-level data, they consist of hundreds of terms, many of which are only used by a handful of proteins. 
Thus, they are impractical to use directly as features. 

Keywords were converted into features through dimensionality reduction. 
Principal component analysis was performed on all keywords present in at least $5\%$ of human proteins.
Technical keywords such as ``3D-Structure'' and ``Reference proteome'' were excluded as not pertaining directly to the protein itself. 
Each protein was then represented by the first 6 components, chosen by a dropoff in explained variance after the first 6 components. 
The most important keywords for these components represented a variety of biological concepts, from functional categories such as ``Tumor suppressor'' and ``Lipid biosynthesis'' to structural features such as ``ANK repeat'' and ``Voltage-gated channel''.
These components only accounted for $42\%$ of variance.
However, given the diversity of keywords, it is unlikely a small number of features could fully represent them. 
Uniprot keywords were retrieved in October 2021. 

\subsection{Metrics}

We used a multiclass F1 score to evaluate predictive performance. 
The F1 score is the harmonic mean of the precision and recall.
In order to extend the F1 score from binary classification to a multiclass setting, the multiclass F1 score is defined as the average F1 score over all classes, treating each class as a one-against-all binary classification. 
This aggressively penalizes predictions if either the precision or recall is low. It is calculated as:

$$F1 = \frac{1}{L} (\sum\limits_{i=1}^{L} \frac{2P_i R_i}{P_i + R_i})$$

where $L$ is the number of class labels, $P_i$ is the precision of class $i$ treated as a one-against-all binary classification, calculated as:

$$P_i = \frac{TP_i}{TP_i + FP_i}$$

and $R_i$ is the recall of class $i$ treated as a one-against-all binary classification, calculated as:

$$R_i = \frac{TP_i}{TP_i + FN_i}$$

where $FP_i$ is the number of instances incorrectly predicted to be class $i$, $TP_i$  is the number of instances correctly predicted to be class $i$, and $FN_i$  is the number of instances incorrectly predicted to not be class $i$. 
This is considered the ``macro'' F1 score. 

\section{Supplementary Results}
\subsection{Database Localization Disagreements}
\begin{figure*}[htbp]
         \centering
         \includegraphics[width=\textwidth]{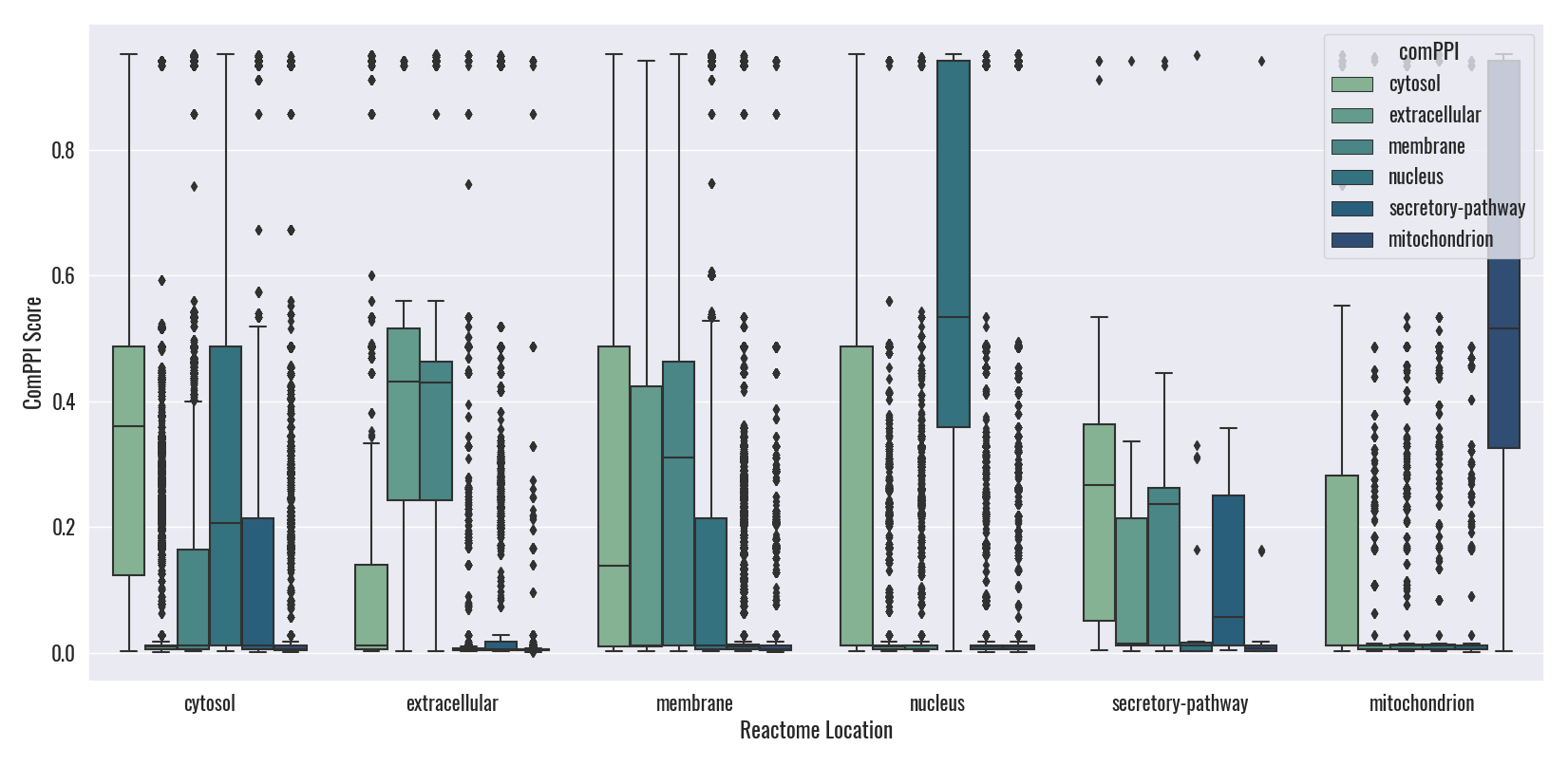}
         \caption{Distribution of ComPPI protein scores by the localization of Reactome edges they belong to. Scores are the probability of a protein being in a given subcellular location.}
         \label{fig:comPPI_scoredist}
     \end{figure*}

\begin{figure*}[htbp]
         \centering
         \includegraphics[width=\textwidth]{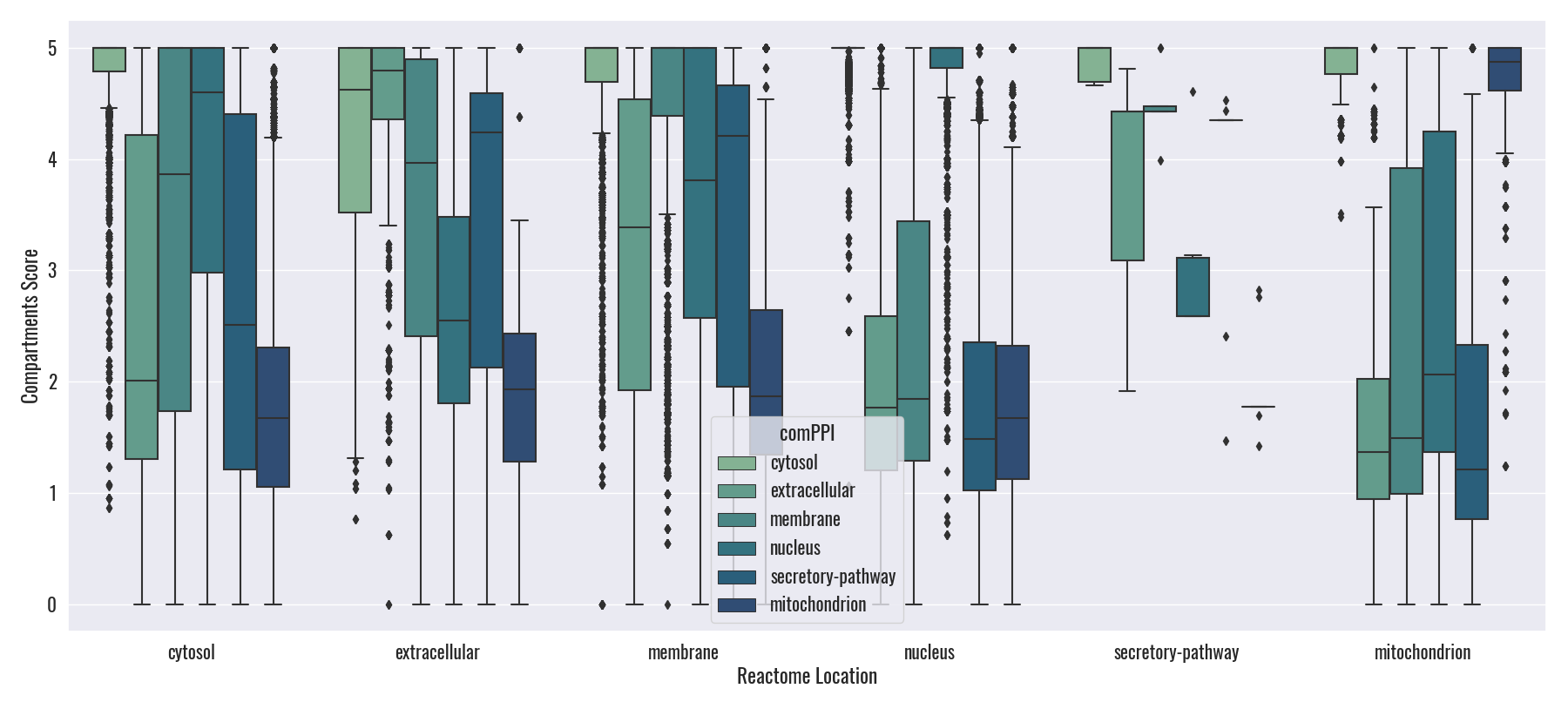}
         \caption{Distribution of Compartments protein scores by the localization of Reactome edges they belong to. Scores are confidence scores of a protein being in a given subcellular location, weighted by the type and amount of evidence available.}
         \label{fig:compartments_scoredist}
     \end{figure*}

The Compartments database has more disagreement with pathway databases than ComPPI. 
Although the range of ComPPI scores for all interaction localizations is wide, including a significant number of proteins existing in localizations where they have a score of 0, in all subcellular locations except for secretory-pathways the median ComPPI score is highest for the corresponding Reactome localization (Figure \ref{fig:comPPI_scoredist}). 
However, in Compartments the cytosol has the highest median score across the majority of all Reactome localizations (Figure \ref{fig:compartments_scoredist}). 

Reactome and PathBank also have some disagreement in their localizations. 
On average, highly matching pathways Reactome and PathBank have $79\%$ agreement in their assigned interaction localizations. 
Highly matching pathways were calculated as those with at least $90\%$ of edges in one pathway appearing in the other. 
This shows a moderate amount of disagreement even between manually curated pathway databases.

\subsection{Pathway Localization Prediction}
\begin{figure*}[htbp]
         \centering
         \includegraphics[width=\textwidth]{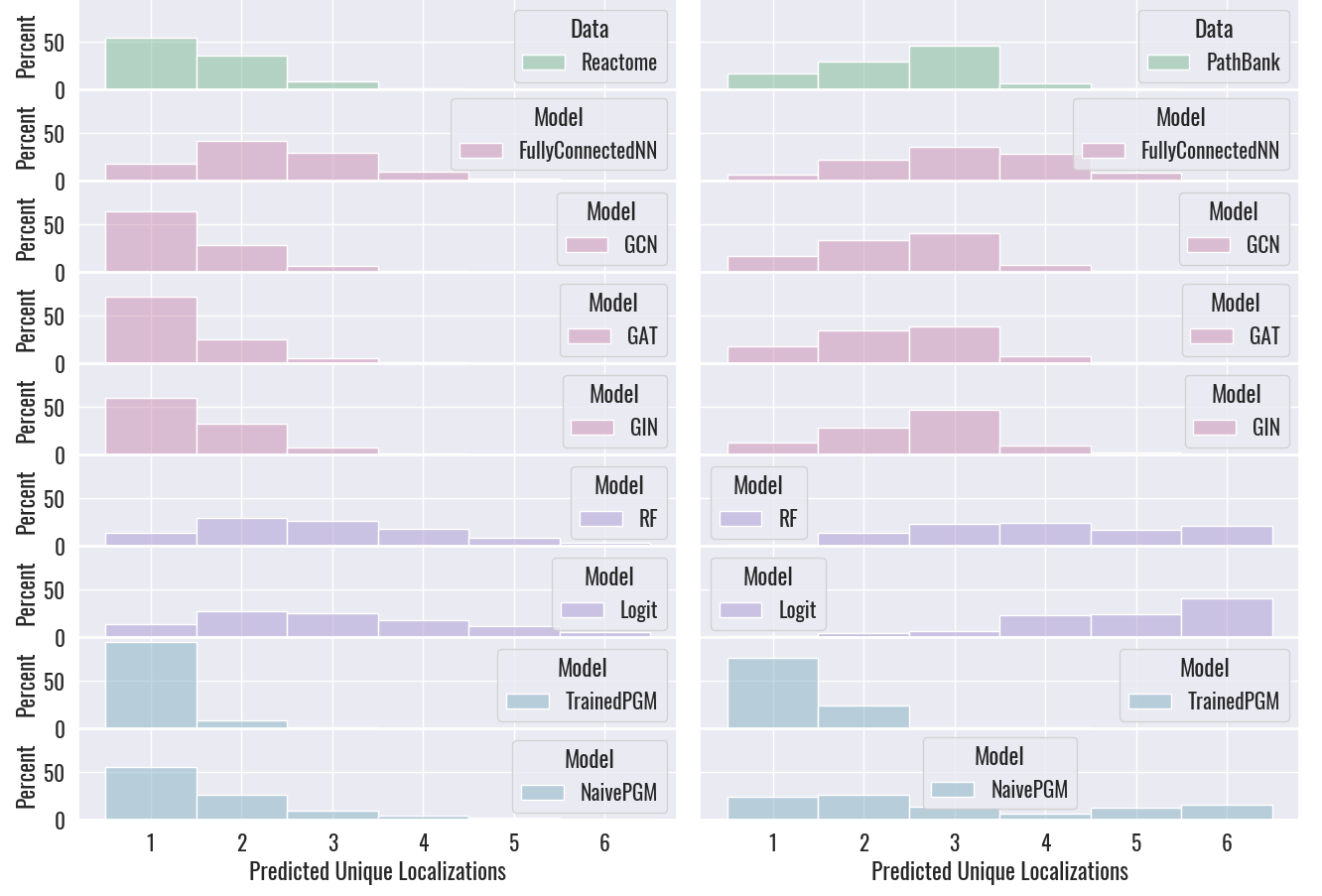}
         \caption{Distributions of the number of unique localizations in each pathway database and predicted by each model on each pathway database. The left column shows distributions for predictions on the Reactome pathway database, and the right column shows distributions for predictions on the PathBank database.}
         \label{fig:unique_dists}
     \end{figure*}

Overall, models without any method to transfer information across a pathway, the RF, Logit, and fully connected neural network models, tended to undersmooth within each pathway (Figure~\ref{fig:unique_dists}). 
The distributions for the RF, Logit, and fully connected neural network models are right-skewed as compared to the true distribution, with a sizable proportion of pathways being predicted to have 5 or 6 different localizations. 
This is unsurprising, as these models contained no topological information with which to encourage proteins belonging to the same pathway to have the same localization. 
These models greatly underestimated the proportion of pathways with a single localization. 

The TrainedPGM in particular tended to oversmooth (Figure~\ref{fig:unique_dists}). 
It predicted almost all pathways as having a single localization across both datasets. 
It also performs particularly poorly for pathways with 2 and 3 unique localizations.
This is likely due to these pathways being more evenly split between multiple localizations than those with 4 or 5 localizations, resulting in a single localization prediction to perform poorly on these pathways. 

\clearpage

\section{Supplementary Figures}
 
    \begin{figure*}[htbp]
         \centering
         \makebox[\textwidth][c]{\includegraphics[width=1.0\textwidth]{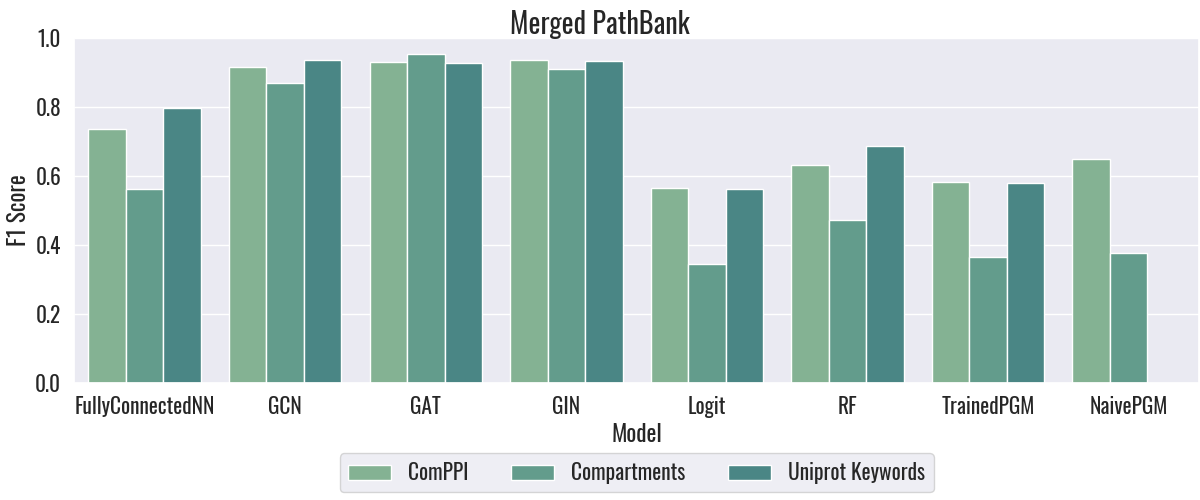}}
         \caption{F1 score of predictive performance on PathBank localizations. All pathway edges are merged and measured together, resulting in 97,792 edges total.}
         \label{fig:merged_pb_f1}
     \end{figure*}

    \begin{figure*}[htbp]
         \centering
         \makebox[\textwidth][c]{\includegraphics[width=1.0\textwidth]{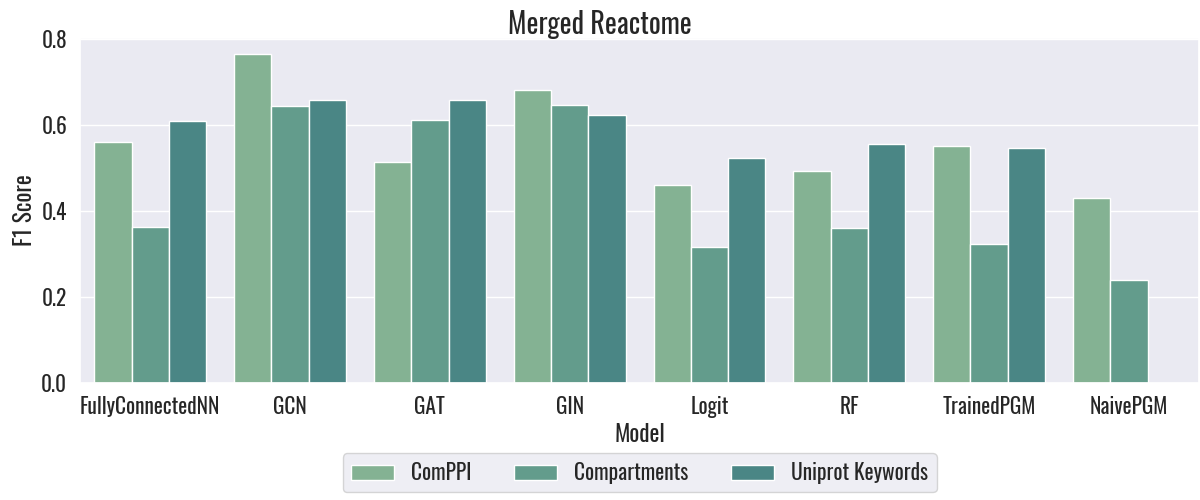}}
         \caption{F1 score of predictive performance on Reactome localizations. All pathway edges are merged and measured together, resulting in 83,855 edges total.}
         \label{fig:merged_reactome_f1}
     \end{figure*}
     
\clearpage

\section{Supplementary Tables}

\begin{table}[h]
\centering
Table S1: Hyperparameter ranges searched for each model. % Still hard-coded
    \caption{Classifier and neural network models and hyperparameters ranges searched for each. Chosen hyperparameter values can be found in Table \ref{tab:params_values}}
    \label{tab:params}
\renewcommand{\arraystretch}{1}
\centering
    {\sscriptsize
        \begin{tabular}{|l|l|l|l|}
            \hline
            Model & Hyperparameter & Description & Range \\ \hline
            \multirow{6}{*}{All Neural Networks} 
                    & \multicolumn{1}{|l|}{Learning Rate} & \multicolumn{1}{|l|}{Learning rate for training.} & \multicolumn{1}{|l|}{$10^{-5}-0.01$} \\\cline{2-4}
                    & \multicolumn{1}{|l|}{Linear Depth} & \multicolumn{1}{|l|}{The number of linear layers.} & \multicolumn{1}{|l|}{$1-5$} \\\cline{2-4}
                    & \multicolumn{1}{|l|}{Convolutional Depth} & \multicolumn{1}{|l|}{The number of convolutional layers.} & \multicolumn{1}{|l|}{$1-10$} \\
                     & \multicolumn{1}{|l|}{} & \multicolumn{1}{|l|}{Not used by linear network.} & \multicolumn{1}{|l|}{} \\\cline{2-4}
                    & \multicolumn{1}{|l|}{Dim} & \multicolumn{1}{|l|}{The number of dimensions in hidden layers.} & \multicolumn{1}{|l|}{$24-128$} \\\cline{2-4}
                    & \multicolumn{1}{|l|}{Dropout} & \multicolumn{1}{|l|}{Whether or not to add dropout while training.} & \multicolumn{1}{|l|}{True, False}\\\hline
            Graph Convolutional Network & & No Unique Parameters. & \\                \hline
            Graph Attention Network & Heads & The number of attention heads. & $1-5$ \\                \hline
            Graph Isomorphism Network &  & No Unique Parameters. & \\                \hline
            Fully Connected Network & Activation & Activation function used. & Tanh, ReLU \\                \hline
            \multirow{4}{*}{Random Forest} 
                    & \multicolumn{1}{|l|}{max\_depth} & \multicolumn{1}{|l|}{Maximum tree depth.} & \multicolumn{1}{|l|}{$1-10$} \\\cline{2-4}
                    & \multicolumn{1}{|l|}{min\_samples\_split} & \multicolumn{1}{|l|}{Minimum samples to create branches.} & \multicolumn{1}{|l|}{$2-10$} \\\cline{2-4}
                    & \multicolumn{1}{|l|}{n\_estimators} & \multicolumn{1}{|l|}{Number of trees.} & \multicolumn{1}{|l|}{$1-100$} \\\cline{2-4}
                    & \multicolumn{1}{|l|}{class\_weight} & \multicolumn{1}{|l|}{Whether to balance class weights.} & \multicolumn{1}{|l|}{True, False} \\\hline
            \multirow{4}{*}{Logistic Regression} 
                    & \multicolumn{1}{|l|}{tol} & \multicolumn{1}{|l|}{Tolerance for training.} & \multicolumn{1}{|l|}{$10^{-6}-0.1$} \\\cline{2-4}
                    & \multicolumn{1}{|l|}{penalty} & \multicolumn{1}{|l|}{Regularization penalty to use.} & \multicolumn{1}{|l|}{L2, None} \\\cline{2-4}
                    & \multicolumn{1}{|l|}{C} & \multicolumn{1}{|l|}{Regularization strength (lower is stronger).} & \multicolumn{1}{|l|}{$0.01-100$} \\\cline{2-4}
                    & \multicolumn{1}{|l|}{class\_weight} & \multicolumn{1}{|l|}{Whether to balance class weights.} & \multicolumn{1}{|l|}{True, False} \\\hline
        \end{tabular}}

\end{table}

\begin{longtblr}[
    caption = {All hyperparameter values used.},
    label = {tab:params_values},
]{
  colspec = {|lll|l|r|}, width = 1.3\linewidth, hlines,
  rowhead = 1, rowfoot = 0, rows={font=\footnotesize}, columns = {30mm}
} 
Model & Dataset & Features & Hyperparameter & Value \\ 
\hline FullyConnectedNN & Reactome & ComPPI & &\\
& & & lRate & $0.002$ \\
& & & l\_depth & $2$ \\
& & & dropout & $0.500$ \\
& & & dim & $83$ \\
& & & activation & tanh \\
\hline FullyConnectedNN & Reactome & Compartments & &\\
& & & lRate & $0.008$ \\
& & & l\_depth & $1$ \\
& & & dropout & $0.500$ \\
& & & dim & $82$ \\
& & & activation & tanh \\
\hline FullyConnectedNN & Reactome & Uniprot KW & &\\
& & & lRate & $8.38e-04$ \\
& & & l\_depth & $3$ \\
& & & dropout & $0.500$ \\
& & & dim & $98$ \\
& & & activation & relu \\
\hline FullyConnectedNN & PathBank & ComPPI & &\\
& & & lRate & $0.010$ \\
& & & l\_depth & $1$ \\
& & & dropout & $0$ \\
& & & dim & $41$ \\
& & & activation & tanh \\
\hline FullyConnectedNN & PathBank & Compartments & &\\
& & & lRate & $0.009$ \\
& & & l\_depth & $2$ \\
& & & dropout & $0$ \\
& & & dim & $103$ \\
& & & activation & relu \\
\hline FullyConnectedNN & PathBank & Uniprot KW & &\\
& & & lRate & $0.006$ \\
& & & l\_depth & $5$ \\
& & & dropout & $0.500$ \\
& & & dim & $81$ \\
& & & activation & tanh \\
\hline GCN & Reactome & ComPPI & &\\
& & & lRate & $0.004$ \\
& & & l\_depth & $1$ \\
& & & dropout & $0$ \\
& & & dim & $81$ \\
& & & c\_depth & $6$ \\
\hline GCN & Reactome & Compartments & &\\
& & & lRate & $1.48e-04$ \\
& & & l\_depth & $1$ \\
& & & dropout & $0.500$ \\
& & & dim & $108$ \\
& & & c\_depth & $2$ \\
\hline GCN & Reactome & Uniprot KW & &\\
& & & lRate & $0.003$ \\
& & & l\_depth & $4$ \\
& & & dropout & $0$ \\
& & & dim & $87$ \\
& & & c\_depth & $3$ \\
\hline GCN & PathBank & ComPPI & &\\
& & & lRate & $0.002$ \\
& & & l\_depth & $1$ \\
& & & dropout & $0$ \\
& & & dim & $44$ \\
& & & c\_depth & $8$ \\
\hline GCN & PathBank & Compartments & &\\
& & & lRate & $0.006$ \\
& & & l\_depth & $2$ \\
& & & dropout & $0$ \\
& & & dim & $96$ \\
& & & c\_depth & $1$ \\
\hline GCN & PathBank & Uniprot KW & &\\
& & & lRate & $0.003$ \\
& & & l\_depth & $3$ \\
& & & dropout & $0$ \\
& & & dim & $108$ \\
& & & c\_depth & $5$ \\
\hline GAT & Reactome & ComPPI & &\\
& & & lRate & $0.003$ \\
& & & l\_depth & $2$ \\
& & & dropout & $0$ \\
& & & dim & $48$ \\
& & & c\_depth & $7$ \\
& & & num\_heads & $4$ \\
\hline GAT & Reactome & Compartments & &\\
& & & lRate & $2.78e-04$ \\
& & & l\_depth & $4$ \\
& & & dropout & $0.500$ \\
& & & dim & $46$ \\
& & & c\_depth & $1$ \\
& & & num\_heads & $5$ \\
\hline GAT & Reactome & Uniprot KW & &\\
& & & lRate & $0.010$ \\
& & & l\_depth & $2$ \\
& & & dropout & $0$ \\
& & & dim & $31$ \\
& & & c\_depth & $4$ \\
& & & num\_heads & $1$ \\
\hline GAT & PathBank & ComPPI & &\\
& & & lRate & $0.003$ \\
& & & l\_depth & $4$ \\
& & & dropout & $0.500$ \\
& & & dim & $45$ \\
& & & c\_depth & $4$ \\
& & & num\_heads & $3$ \\
\hline GAT & PathBank & Compartments & &\\
& & & lRate & $0.002$ \\
& & & l\_depth & $3$ \\
& & & dropout & $0.500$ \\
& & & dim & $43$ \\
& & & c\_depth & $3$ \\
& & & num\_heads & $4$ \\
\hline GAT & PathBank & Uniprot KW & &\\
& & & lRate & $0.004$ \\
& & & l\_depth & $1$ \\
& & & dropout & $0$ \\
& & & dim & $39$ \\
& & & c\_depth & $2$ \\
& & & num\_heads & $4$ \\
\hline GIN & Reactome & ComPPI & &\\
& & & lRate & $0.005$ \\
& & & l\_depth & $2$ \\
& & & dropout & $0$ \\
& & & dim & $95$ \\
& & & c\_depth & $3$ \\
\hline GIN & Reactome & Compartments & &\\
& & & lRate & $9.77e-04$ \\
& & & l\_depth & $4$ \\
& & & dropout & $0.500$ \\
& & & dim & $24$ \\
& & & c\_depth & $1$ \\
\hline GIN & Reactome & Uniprot KW & &\\
& & & lRate & $0.001$ \\
& & & l\_depth & $3$ \\
& & & dropout & $0.500$ \\
& & & dim & $60$ \\
& & & c\_depth & $2$ \\
\hline GIN & PathBank & ComPPI & &\\
& & & lRate & $0.002$ \\
& & & l\_depth & $4$ \\
& & & dropout & $0$ \\
& & & dim & $68$ \\
& & & c\_depth & $1$ \\
\hline GIN & PathBank & Compartments & &\\
& & & lRate & $0.001$ \\
& & & l\_depth & $3$ \\
& & & dropout & $0.500$ \\
& & & dim & $102$ \\
& & & c\_depth & $3$ \\
\hline GIN & PathBank & Uniprot KW & &\\
& & & lRate & $9.86e-04$ \\
& & & l\_depth & $2$ \\
& & & dropout & $0$ \\
& & & dim & $80$ \\
& & & c\_depth & $1$ \\
\hline Logit & Reactome & ComPPI & &\\
& & & C & $0.620$ \\
& & & class\_weight & balanced \\
& & & penalty & l2 \\
& & & tol & $1.00e-06$ \\
\hline Logit & Reactome & Compartments & &\\
& & & C & $91.893$ \\
& & & class\_weight & balanced \\
& & & penalty & l2 \\
& & & tol & $0.062$ \\
\hline Logit & Reactome & Uniprot KW & &\\
& & & C & $7.604$ \\
& & & class\_weight & balanced \\
& & & penalty & l2 \\
& & & tol & $0.078$ \\
\hline Logit & PathBank & ComPPI & &\\
& & & C & $0.044$ \\
& & & class\_weight & balanced \\
& & & penalty & l2 \\
& & & tol & $1.84e-05$ \\
\hline Logit & PathBank & Compartments & &\\
& & & C & $0.451$ \\
& & & class\_weight & balanced \\
& & & penalty & l2 \\
& & & tol & $4.71e-06$ \\
\hline Logit & PathBank & Uniprot KW & &\\
& & & C & $33.176$ \\
& & & class\_weight & balanced \\
& & & penalty & l2 \\
& & & tol & $0.033$ \\
\hline RF & Reactome & ComPPI & &\\
& & & class\_weight & balanced \\
& & & max\_depth & $3$ \\
& & & min\_samples\_split & $2$ \\
& & & n\_estimators & $76$ \\
\hline RF & Reactome & Compartments & &\\
& & & class\_weight & balanced \\
& & & max\_depth & $9$ \\
& & & min\_samples\_split & $3$ \\
& & & n\_estimators & $81$ \\
\hline RF & Reactome & Uniprot KW & &\\
& & & class\_weight & balanced \\
& & & max\_depth & $6$ \\
& & & min\_samples\_split & $3$ \\
& & & n\_estimators & $100$ \\
\hline RF & PathBank & ComPPI & &\\
& & & class\_weight & balanced \\
& & & max\_depth & $10$ \\
& & & min\_samples\_split & $10$ \\
& & & n\_estimators & $72$ \\
\hline RF & PathBank & Compartments & &\\
& & & class\_weight & balanced \\
& & & max\_depth & $10$ \\
& & & min\_samples\_split & $10$ \\
& & & n\_estimators & $92$ \\
\hline RF & PathBank & Uniprot KW & &\\
& & & class\_weight & balanced \\
& & & max\_depth & $10$ \\
& & & min\_samples\_split & $10$ \\
& & & n\_estimators & $58$ \\
\end{longtblr}

\end{document}